\documentclass[journal,comsoc]{IEEEtran}
\usepackage{hyperref}
\hypersetup{hypertex=true,
colorlinks=true,
linkcolor=blue,
anchorcolor=blue,
citecolor=blue}

\usepackage{booktabs}
\usepackage{makecell}
\IEEEoverridecommandlockouts
\usepackage{cite}
\usepackage{amsmath,amssymb,amsfonts}
\usepackage{algorithmic}
\usepackage{algorithm}
\usepackage{graphicx}
\usepackage{textcomp}
\usepackage{xcolor}
\usepackage{geometry}
\usepackage[caption=false,font=footnotesize,
labelfont=rm,textfont=rm]{subfig}
\renewcommand{\baselinestretch}{1}
\def\BibTeX{{\rm B\kern-.05em{\sc i\kern-.025em b}\kern-.08em
    T\kern-.1667em\lower.7ex\hbox{E}\kern-.125emX}}
\begin{document}
\title{UAV-Enabled Passive 6D Movable Antennas: Joint Deployment and Beamforming Optimization
}

\author{Changhao Liu, Weidong Mei, \IEEEmembership{Member, IEEE}, Peilan Wang, Yinuo Meng, Boyu Ning, \IEEEmembership{Member, IEEE}, and Zhi Chen, \IEEEmembership{Senior Member, IEEE }
\thanks{Part of this paper will be presented
at the IEEE Global Communications Conference (Globecom), Cape Town, South Africa, 2024.\cite{Globecom}

C. Liu, W. Mei, P. Wang, B. Ning, and Z. Chen are with the National Key Laboratory of Wireless Communications, University of Electronic Science and Technology of China, Chengdu 611731, China (e-mail:  lch3001@163.com, wmei@uestc.edu.cn, peilan\_wangle@uestc.edu.cn, boydning@outlook.com, chenzhi@uestc.edu.cn).

Y. Meng is with the Glasgow College, University of Electronic Science and Technology of China, Chengdu 611731, China (e-mail: 2022190502030@std.uestc.edu.cn).}}

\maketitle

\begin{abstract}
Intelligent reflecting surface (IRS) is composed of numerous passive reflecting elements and can be mounted on unmanned aerial vehicles (UAVs) to achieve six-dimensional (6D) movement by adjusting the UAV's three-dimensional (3D) location and 3D orientation simultaneously. Hence, in this paper, we investigate a new UAV-enabled passive 6D movable antenna (6DMA) architecture by mounting an IRS on a UAV and address the associated joint deployment and beamforming optimization problem. In particular, we consider a passive 6DMA-aided multicast system with a multi-antenna base station (BS) and multiple remote users, aiming to jointly optimize the IRS's location and 3D orientation, as well as its passive beamforming to maximize the minimum
received signal-to-noise ratio (SNR) among all users under the practical angle-dependent signal reflection model. However, this optimization problem is challenging to be optimally solved due to the intricate relationship between the users' SNRs and the IRS's location and orientation.
To tackle this challenge, we first focus on a simplified case with a single user, showing that one-dimensional (1D) orientation suffices to achieve the optimal performance. Next, we show that for any given IRS's location, the optimal 1D orientation can be derived in closed form, based on which several useful insights are drawn. 
To solve the max-min SNR problem in the general multi-user case, we propose an alternating optimization (AO) algorithm by alternately optimizing the IRS's beamforming and location/orientation via successive convex approximation (SCA) and hybrid coarse- and fine-grained search, respectively. To avoid undesirable local sub-optimal solutions, a Gibbs sampling (GS) method is proposed to generate new IRS locations and orientations for exploration in each AO iteration.
Numerical results validate our theoretical analyses and demonstrate the superiority of our proposed AO algorithm with GS to conventional AO and other baseline deployment strategies with location or orientation optimization only.
\end{abstract}


\section{Introduction}
Intelligent reflecting surface (IRS), also known as  reconfigurable intelligent surface (RIS), has recently emerged as a promising technology to enhance the performance of  wireless communication systems in an energy-efficient and cost-effective manner. Specifically, an IRS is a planar meta-surface consisting of a large number of sub-wavelength-size passive reflecting elements, each of which is capable of reflecting the impinging signals with an adjustable phase shift. By jointly optimizing the phase shifts of its reflecting elements (i.e., passive beamforming), the IRS can alter the direction of its reflected signals, thereby realizing various purposes such as coverage extension\cite{IRS1[4],IRS1[3],IRS1[1],IRS1[2],IRS-TUTOR1, IRS-TUTOR3}, interference suppression\cite{IRS1[1],IRS1[2],IRS1[3],IRS-TUTOR1,IRS-TUTOR3}, wireless power transfer\cite{IRS1[1], IRS1[2], IRS-TUTOR1, IRS-TUTOR2, IRS-TUTOR3}, target sensing\cite{IRS1[1], IRS1[2], IRS-TUTOR1}, and so on.

\begin{figure}[tb]
\centerline{\includegraphics[width=0.43\textwidth]{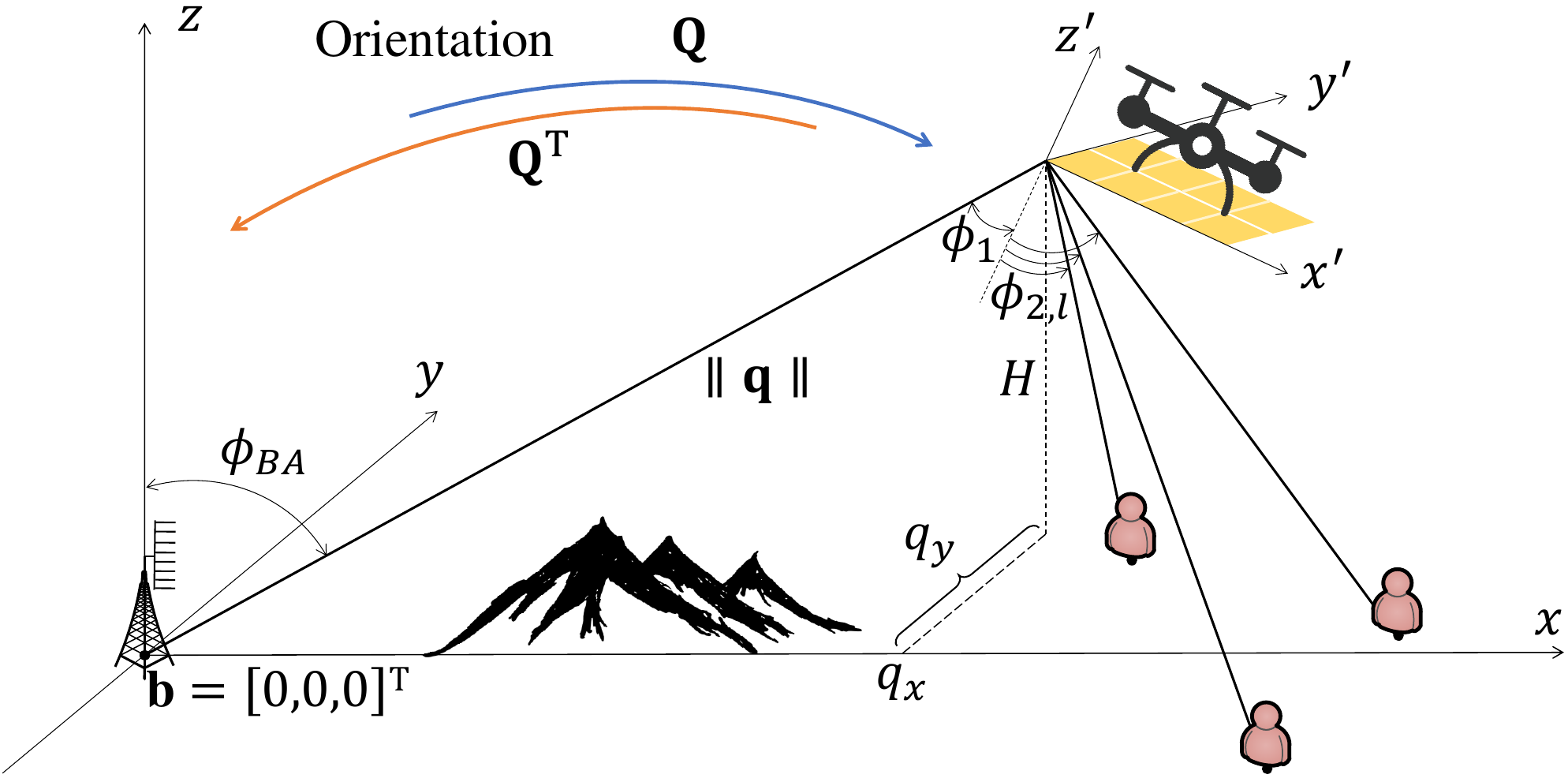}}
\caption{A UAV-enabled passive 6DMA-assisted multicast system.}
\label{sysmod}
\end{figure}

However, the deployment of the IRS plays a significant role in realizing the above benefits, as an IRS can only reflect signals from/to its pointing half space and may cause severe signal power loss due to its passive reflection. Hence, some prior works have delved into the IRS deployment optimization under different system setups\cite{IRS1[8],IRS1,FUMIN,IRS1[9], IRS1[10], TDB2}.
For example, the authors in \cite{IRS1[8]} compared the capacity regions achievable by two IRS deployment strategies with the IRS/IRSs deployed near the base station (BS) and each of distributed users, respectively, and showed the superiority of the former over the latter under the same total number of reflecting elements. 
In \cite{IRS1} and \cite{FUMIN}, the authors optimized IRS's deployment to achieve the trade-off between the indoor coverage performance and total deployment cost using graph-based approaches.
A two-timescale optimization framework was proposed in \cite{IRS1[9]}, where the IRS's deployment and passive beamforming were optimized based on the long- and short-term channel knowledge, respectively. 
In \cite{IRS1[10]}, the authors jointly optimized the beamforming and deployment of an IRS to maximize the non-outage secrecy rate in a secure wireless communication system. 
In \cite{TDB2}, the authors optimized the deployment and beamforming of two IRSs distributed on the same wall to extend the BS's signal coverage within a target region.
However, all of the above works \cite{IRS1[8],IRS1,FUMIN,IRS1[9], IRS1[10], TDB2} assumed terrestrial IRSs, which still face limitations due to their generally fixed positions and $180^\circ$ half-space reflection, potentially failing to cover all BSs/users.

To overcome the above issues, some previous studies have proposed mounting the IRS on an aerial platform (e.g., an unmanned aerial vehicle (UAV)), referred to as aerial IRS (AIRS)\cite{IRSAAV2,AIRS,AIRS3,AIRS-1}. Compared to its terrestrial counterpart, an AIRS is more likely to achieve $360^\circ$ panoramic full-angle reflection, due to its much higher altitude than the terrestrial BSs and users \cite{IRSAAV2}. In addition, its location and even orientation can be flexibly adjusted by leveraging the maneuverability of UAVs. Motivated by the above benefits of the AIRS, the authors in \cite{AIRS}  and \cite{AIRS3} aimed to jointly optimize the location and reflection of a passive AIRS serving multiple users.
However, both of them assumed isotropic signal reflection by each AIRS reflecting element, while the AIRS's passive beamforming gain is practically dependent on its incident and reflected angles \cite{AIRS-1}. In view of this fact, the orientation of the AIRS can be further optimized to enhance the communication performance. In \cite{ORI1} and \cite{ORI2}, the authors have investigated the joint location and orientation optimization for a terrestrial IRS and an AIRS, respectively. Nevertheless, they only considered one-dimensional (1D) orientation adjustment, despite the additional degrees of freedom (DoFs) offered by three-dimensional (3D) orientation/posture control. Moreover, both \cite{ORI1} and \cite{ORI2} considered a single-user setup, while the effects of AIRS orientation on the multi-user communication system remains unknown.

To fully exploit all six-dimensional (6D) DoFs available for UAV-mounted IRSs or AIRSs, i.e., 3D location and 3D orientation, we propose a new architecture of {\it UAV-enabled passive 6D movable antenna (6DMA)} in this paper, as shown in Fig.~\ref{sysmod}. Note that compared to the terrestrial 6DMA proposed in \cite{QB} and \cite{SX1}, our proposed UAV-enabled 6DMA offers a broader range of tuning for antenna positioning and orientation thanks to the mobility of the UAV. Focusing on a multicast communication system, we aim to jointly optimize the AIRS's location and 3D orientation, as well as its passive beamforming to maximize the minimum received signal-to-noise ratio (SNR) among all users under the practical angle-dependent signal reflection model. To the best of our knowledge, this is the first work investigating the performance optimization of a UAV-enabled passive 6DMA in the literature. Our main contributions are summarized as follows.
\begin{itemize}
\item To gain useful insights into the passive 6DMA, we first solve the SNR maximization problem in a simplified single-user setup. Our analysis demonstrates that in this case, 1D orientation is sufficient to achieve the maximum received SNR at the user. Next, we show that for any given AIRS location, the optimal 1D orientation can be derived in closed form, based on which several useful insights are drawn. Furthermore, in some special cases regarding the AIRS's altitude and the BS-user distance, we also derive the AIRS's optimal location in closed form.
\item However, the max-min SNR optimization problem in the general multi-user case is much more challenging to be optimally solved due to the intricate coupling of the AIRS's 6D movement and beamforming. To tackle this challenge, we propose an alternating optimization (AO) algorithm by alternately optimizing the AIRS's passive beamforming and location/orientation via successive convex approximation (SCA) and multi-dimensional search, respectively. Particularly, to reduce the searching complexity in location/orientation optimization with given AIRS passive beamforming, we first conduct a coarse-grained search to quickly obtain a locally sub-optimal solution, followed by a fine-grained search near this solution. Furthermore, to avoid undesirable trapping at sub-optimal solutions in the conventional AO, a Gibbs sampling (GS) method is proposed to generate a sequence of samples of the AIRS's candidate locations and orientations via a probability-based Markov chain for solution exploration. 
\item Numerical results validate our theoretical analyses and demonstrate the superiority of our proposed enhanced AO algorithm with GS to conventional AO without GS and other baseline deployment strategies with either location or orientation optimization only and that without accounting for the AIRS's angle-dependent signal reflection. 
It is also shown that the proposed algorithm shows varying characteristics in balancing the passive beamforming gain, end-to-end path loss, and effective aperture gain among the users, depending on their dense or sparse geographic distributions. 
\end{itemize}
It is worth noting that the relative locations of all reflecting elements of the AIRS keep unchanged in our proposed passive 6DMA, while they may also be altered to bring even more DoFs for performance enhancement \cite{QA,QC,QD,QE}.

The rest of this paper is organized as follows. Section II introduces the system model for the UAV-enabled passive 6DMA-aided multicast system and formulates the design problem. In Section III, we consider a simplified single-user case and show the optimality of 1D orientation, deriving its optimal solution in closed form. In Section IV, we propose an enhanced AO algorithm with GS to solve the max-min SNR problem in the general multi-user setup. Section V presents numerical results to evaluate the performance of our proposed algorithm. Section VI concludes the paper. 

{\it Notations:}  For a complex number $s$, symbols $\angle s$, $\lvert s \rvert$, and $s^*$ denote its phase, amplitude, and conjugate, respectively. For a vector $\mathbf{x}$, symbols $\mathbf{x}^T$, $\mathbf{x}^H$, $\|\mathbf{x}\|$, $(\mathbf{x})_n$,  and $\mathrm{diag}(\mathbf{x})$ denote its transpose, conjugate transpose, Euclidean norm, the $n$-th entry, and a diagonal matrix with its entries on the main diagonal, respectively. 
For a matrix $\mathbf{X}$, $\mathbf{X}[m,n]$ denotes the element on the $m$-th row and $n$-th column of $\mathbf{X}$. Symbol $\mathbb{C}^{M\times N}$ denotes the set of $M \times N$ complex-valued matrices. 
For two sets $\mathcal{A}$ and $\mathcal{B}$, $\mathcal{A}\cup \mathcal{B}$ denotes the union of $\mathcal{A}$ and $\mathcal{B}$, and $\mathcal{A} \setminus \mathcal{B}$ denotes the set of elements that belong to $\mathcal{A}$ but are not in $\mathcal{B}$.
The symbols $\emptyset$, $\otimes$, $\odot$, and $j$ denote the empty set, Kronecker product, Hadamard product, and the imaginary unit with $j^2 = -1$, respectively.

\section{System Model and Problem Formulation}
\subsection{Signal Model}
As shown in Fig.~\ref{sysmod}, we consider the downlink transmission from a BS to $N_u$ remote users aided by an AIRS. The direct links between the BS and users are assumed to be non-existent due to the severe path loss caused by large distances. Thanks to UAV's flexible deployment and posture control as well as IRS's beamforming design, the phase shifts and location/orientation of the AIRS can be adaptively tuned to enhance the communication performance based on users' locations. The BS is assumed to be equipped with a uniform linear array (ULA)  with $M$ antennas, while each user is equipped with a single antenna, and the users are distributed in a remote area $\mathcal{A}$. 
In this paper, we focus on a coherent time slot to investigate the performance of our proposed algorithm, during which the AIRS's location and orientation can be viewed as being constant.  

For convenience, we establish a global Cartesian coordinate system (CCS) in the considered system, assuming that the BS is located at the origin, i.e., $\mathbf{b}=[0,0,0]^T$. Let the coordinate of the $l$-th user be $\mathbf{w}_l=[w_{lx},w_{ly},0]^T$, $l\in {\cal N}_u \triangleq \{1,2,\cdots,N_u\}$.
The ULA at the BS is assumed to be parallel to the $z$-axis in the global CCS. 
In the presence of the UAV-enabled orientation of the IRS, to ease the computation of the angle information from the AIRS to the BS/users, we also define a local CCS at the AIRS which lies in its $x'$-$y'$ plane, as shown in Fig.~\ref{sysmod}. 
Without loss of generality, we select the bottom-left element of the AIRS as its reference element and denote its coordinate by $\mathbf{q}=[q_x, q_y, H]^T$, where $q_x$ and $q_y$ denote its projection onto the $x$- and $y$-axis, respectively, with $H$ denoting its altitude. 
 The AIRS is assumed to be equipped with a uniform planar array (UPA) with $N=N_x\times N_y$ reflecting elements, where $N_x$ and $N_y$ denote the numbers of reflecting elements along the $x'$- and $y'$-axes of the local CCS, respectively. The distances between any two adjacent antennas and elements at the BS and the AIRS are denoted as $d_{tx}$ and $d_{rs}$, respectively.

The 3D orientation of the AIRS can be represented by $\boldsymbol{\psi} = [\psi_z,\psi_y,\psi_x]^T$, where $\psi_z$, $\psi_y$, and $\psi_x$ are Euler angles denoting the AIRS's degree of orientation around $z'$-, $y'$-, and $x'$-axis, respectively. Then, the relationship between the global CCS and the local CCS is characterized by the following orientation matrix \cite{AIRS-2},
\begin{equation} \label{Q}
\mathbf{Q}(\boldsymbol{\psi})=\mathbf{Q}_z(\psi_z)\,\mathbf{Q}_y(\psi_y)\,\mathbf{Q}_x(\psi_x),
\end{equation}
where $\mathbf{Q}_z(\psi_z)$ indicates the orientation of $\psi_z$ radians around the $z$-axis and is given by
\begin{equation}{
\mathbf{Q}_z(\psi_z) = \begin{bmatrix}
 \cos\psi_z & -\sin\psi_z & 0\\
 \sin\psi_z & \cos\psi_z & 0\\
  0&  0&1
\end{bmatrix},}
\end{equation}
$\mathbf{Q}_y(\psi_y)$ indicates the orientation of $\psi_y$ radians around the $y$-axis and is given by
\begin{equation}
\mathbf{Q}_y(\psi_y) = \begin{bmatrix}
 \cos\psi_y & 0 & \sin\psi_y\\
 0 & 1 & 0\\
  -\sin\psi_y&  0&\cos\psi_y
\end{bmatrix},
\end{equation}
and $\mathbf{Q}_x(\psi_x)$ indicates the orientation of $\psi_x$ radians around the $x$-axis and is given by
\begin{equation}
\mathbf{Q}_x(\psi_x) = \begin{bmatrix}
1 & 0 & 0\\
0 & \cos\psi_x & -\sin\psi_x\\
0&  \sin\psi_x&\cos\psi_x
\end{bmatrix}.
\end{equation}
Based on \eqref{Q}, for any given 3D location $\mathbf{p}$ in the global CCS, its corresponding coordinates in the local CCS are given by \cite{AIRS-2}
\begin{equation}
\mathbf{p}^{\text{local}} = \mathbf{Q}^T(\boldsymbol{\psi})(\mathbf{p-q}).
\end{equation}
Obviously, we have $\mathbf{q}^{\text{local}}=[0,0,0]^T$, i.e., the reference element of the AIRS is at the origin of the local CCS. Moreover, the coordinate of the BS in the local CCS are given by
\begin{equation}	\label{txlocal}
\mathbf{b}^{\text{local}} = \mathbf{Q}^T(\boldsymbol{\psi})(\mathbf{b}-\mathbf{q}) =[b_{x}^{\text{local}},b_{y}^{\text{local}},b_{z}^{\text{local}}]^T,
\end{equation}
and those of the users in the local CCS are 
\begin{equation}	\label{rxlocal}
\mathbf{w}_l^{\text{local}} = \mathbf{Q}^T(\boldsymbol{\psi})(\mathbf{w}_l-\mathbf{q})=[w_{lx}^{\text{local}},w_{ly}^{\text{local}}, w_{lz}^{\text{local}}]^T\!,\; l \in {\cal N}_u.
\end{equation}

Given the above local coordinates, we can define the  angle of departure (AoD) from the BS to the AIRS, the elevation/azimuth angle of arrival (AoA) at the AIRS from the BS, and the elevation/azimuth AoD from the AIRS to user $l$ as $\phi_{BA}$, $\vartheta_{AB}^{(e)}$, $\vartheta_{AB}^{(a)}$, $\phi_{AU\!,\,l}^{(e)}$, and $\phi_{AU\!,\,l}^{(a)}$,  respectively, which can be obtained based on geometry as
\begin{align}
&\;\;\;\;\;\;\;\;\;\;\;\;\;\;\;\;\phi_{BA} =\arccos\frac{H} { \|\mathbf{q}\|},\\
\vartheta_{AB}^{(e)}&=\arccos\frac{-b_z^{\text{local}}}{\|\mathbf{q}\|}, \;\;\;\;\;\;
\vartheta_{AB}^{(a)}=\arctan\frac{b_y^{\text{local}}}{b_x^{\text{local}}}, \\
\phi_{AU\!,\,l}^{(e)}&=\arccos\frac{w_{lz}^{\text{local}}} {\| \mathbf{w}_l-\mathbf{q}\|}, \;
\phi_{AU\!,\,l}^{(a)}=\arctan \frac{w_{ly}^{\text{local}}}{w_{lx}^{\text{local}}}, \; l \in {\cal N}_u.
\end{align} 

Unlike the isotropic signal reflection assumed in \cite{AIRS}, we consider a more practical angle-dependent signal reflection model by taking into account the effective AIRS reflection aperture. To this end, we define $\phi_1$ ($\phi_{2,l}$) as the incident angle (reflection angles) of the BS's signal at the AIRS with respect to (w.r.t.) user $l,\; l \in {\cal N}_u$.
Based on the above local coordinates, it is seen that
\begin{align}	\label{phi1}
\phi_1 &=\arccos \frac{-b_{z}^{\text{local}}}{\|\mathbf{q}\|}=\vartheta_{AB}^{(e)},\\
\phi_{2,l} &=\arccos \frac{-w_{lz}^{\text{local}}}{\|\mathbf{q-w}\|}=\pi-\phi_{AU\!,\,l}^{(e)}. \label{phi2}
\end{align}
Notably, it must hold that $\phi_1 \in [0,\frac{\pi}{2}]$ and $\phi_{2,l} \in [0,\frac{\pi}{2}], \;\forall l \in {\cal N}_u$ to ensure that all users and the BS are located in the reflection space of the AIRS.
By substituting \eqref{Q}-\eqref{rxlocal} into \eqref{phi1} and \eqref{phi2}, we have
\begin{align}	\label{cosphi1}
\cos\phi_1 &= \frac{q_xL_1+q_yL_2+HL_3}{\sqrt{q_x^2+q_y^2+H^2}}, \\
\cos\phi_{2,l} &= \frac{(q_x\!-\!w_{lx})L_1+(q_y\!-\!w_{ly})L_2+HL_3}{\sqrt{(q_x-w_{lx})^2+(q_y-w_{ly})^2+H^2}}, \label{cosphi2}
\end{align} 
where 
\begin{align}
L_1&=\cos\psi_z \sin\psi_y \cos\psi_x + \sin\psi_z\sin\psi_x, \label{L1} \\
L_2&=\sin\psi_z\sin\psi_y\cos\psi_x - 
\cos\psi_z\sin\psi_x, \label{L2} \\
L_3&=\cos\psi_y\cos\psi_x. \label{L3}
\end{align} 

As such, the effective aperture gain due to the AIRS's orientation w.r.t. user $l$ can be expressed as \cite{AIRS-1}
\begin{equation}	\label{F}
F_{\text{AG},l}(\mathbf{q}, \boldsymbol{\psi}) = \cos\phi_1 \cos\phi_{2,l},\; l \in {\cal N}_u.
\end{equation}
It is noted that in the conventional isotropic reflection model, we have $F_{\text{AG},l}=1$ regardless of $\phi_1$ and $\phi_{2,l}$. However, when $\phi_1$ and/or $\phi_{2,l}$ are close to $\frac{\pi}{2}$, the effective aperture gain in \eqref{F} approaches zero.

In this paper, we assume that the UAV/AIRS's altitude, $H$, is fixed at the minimum altitude satisfying free-space LoS propagation from the UAV to the BS/users (e.g., $H \ge 100$ meter (m) in the urban macro scenario \cite{A}) to reduce the end-to-end path loss. In Section V, we will also evaluate the effects of the AIRS's altitude on the overall performance via simulation. As such, the path gain from the BS to the AIRS can be expressed as
$ \beta_1 = \frac{\beta_0}{\|\mathbf{q}\|^2}$, and that from the AIRS to the user $l$ is expressed as $\beta_{2,l} = \frac{\beta_0}{\|\mathbf{q}-\mathbf{w}_l\|^2},\;l \in \mathcal{N}_u$, where $\beta_0$  denotes the path gain at the reference distance of $1$ m. Moreover, due to the practically high altitude of the UAV, we assume far-field propagation between the AIRS and the BS/users. 

Hence, the channel from the BS to the AIRS is given by
\begin{equation}	\label{G}
\mathbf{H}_{BA}  = \sqrt{\beta_1}e^{-j\frac{2\pi \|\mathbf{q}\|}{\lambda}}\mathfrak{\mathbf{a} }_I (\vartheta_{AB}^{(e)},\vartheta_{AB}^{(a)})\mathfrak{\mathbf{a} }^H_B({\phi}_{BA}),
\end{equation}
where $\mathfrak{\mathbf{a}}_I(\vartheta_{AB}^{(e)},\vartheta_{AB}^{(a)})$ and $\mathfrak{\mathbf{a}}_B(\phi_{BA})$ represent the receive and transmit array response vectors at the AIRS and the BS, respectively, which can be expressed as 
\begin{align}	
\mathfrak{\mathbf{a}} _{I}(\vartheta_{AB}^{(e)},\vartheta_{AB}^{(a)})&=\mathfrak{\mathbf{a}} _{Ix}\otimes\mathfrak{\mathbf{a}} _{Iy}\notag\\
=[1,&\cdots,e^{-j\frac{2\pi(N_x-1)}{\lambda}d_{rs}\sin (\vartheta_{AB}^{(e)})\cos(\vartheta_{AB}^{(a)})}]^T  \notag \\
\otimes [1,&\cdots,e^{-j\frac{2\pi(N_y-1)}{\lambda}d_{rs}\sin(\vartheta_{AB}^{(e)})\sin(\vartheta_{AB}^{(a)})}]^T, \label{a1}
\end{align}
and
\begin{equation}	\label{aD}
\mathfrak{\mathbf{a}}_{B}(\phi_{B\!A})\!=\![1,e^{-\!j\!\frac{2\pi}{\lambda}d_{tx}\!\cos\!\phi_{B\!A}}\!,\cdots\!,e^{-\!j\!\frac{2\pi(M-1)}{\lambda}d_{tx}\!\cos\!\phi_{B\!A}}]^T\!.
\end{equation}
Similarly, the channel from the AIRS to user $l$ can be given as 
\begin{equation}	\label{h}
\mathbf{h}_{AU\!,\,l}^H = \sqrt{\beta_{2,l\,}} e^{-j\frac{2\pi \|\mathbf{q}-\mathbf{w}_l\|}{\lambda}}\mathfrak{\mathbf{a} } ^H_2(\phi_{AU\!,\,l}^{(e)}, \phi_{AU\!,\,l}^{(a)}),\;l \in \mathcal{N}_u, 
\end{equation}
where $\mathfrak{\mathbf{a}}_2(\phi_{AU\!,\,l}^{(e)}, \phi_{AU\!,\,l}^{(a)})$ is the transmit array response vector at the AIRS and can be expressed as 
\begin{align}	\label{a2}
\mathfrak{\mathbf{a}} _2(\phi_{AU\!,\,l}^{(e)}, \phi_{AU\!,\,l}^{(a)})&=\mathfrak{\mathbf{a}} _{2x,l}\otimes\mathfrak{\mathbf{a}} _{2y,l} \notag \\
=[1,&\cdots,e^{-j\frac{2\pi(\!N_x-1)}{\lambda}d_{r\!s}\sin(\phi_{AU\!,\,l}^{(e)})\cos(\phi_{AU\!,\,l}^{(a)})}]^T   \notag \\
\otimes [1,&\cdots,e^{-j\frac{2\pi(\!N_y-1)}{\lambda}d_{r\!s}\sin(\phi_{AU\!,\,l}^{(e)})\sin(\phi_{AU\!,\,l}^{(a)})}]^T.
\end{align}

The received signal at user $l$ can be expressed as
\begin{equation}	\label{y}
y_l = \mathrm{\mathbf{h}}_{AU\!,\,l} ^H\mathbf{\Theta}\mathrm{\mathbf{H}}_{BA}\mathbf{v}\sqrt{PF_{\text{AG},l}(\mathbf{q},\boldsymbol{\psi})\,}s + n_w, \;l \in \mathcal{N}_u, 
\end{equation}
where $\mathbf{\Theta} = \mathrm{diag}(e^{j\theta_1}, \cdots, e^{j\theta_N})$ denotes the reflection matrix of the AIRS with $\theta_n$ denoting the phase shift of the $n$-th reflecting element, $\mathbf{v} \in \mathbb{C}^{M \times 1}$ is the transmit beamforming vector of the BS with unit-norm, $P$ and $s$ denote the BS's transmit power and  symbol, respectively, and $n_w \sim {\cal{CN}}(0, \sigma^2)$ is the additive white Gaussian noise (AWGN) with $\sigma^2$ denoting the noise power.
Thus, the received SNR of user $l$ is 
\begin{align}	
\gamma_l(\mathbf{q},\;&\boldsymbol{\psi},\mathbf{\Theta}, \mathbf{v} ) 
=\frac{PF_{\text{AG},l}(\mathbf{q},\boldsymbol{\psi})\lvert\mathrm{\mathbf{h}}_{AU\!,\,l}^H\mathbf{\Theta}\mathrm{\mathbf{H}}_{BA}\mathbf{v}\rvert^2 }{\sigma^2} \notag \\
=&\frac{\bar{P}\beta_0^2F_{\text{AG},l}(\mathbf{q},\boldsymbol{\psi})G_{\text{BF},l}(\mathbf{q},\boldsymbol{\psi},\boldsymbol{\theta})\Big\lvert\mathfrak{\mathbf{a}}_{B}^{H}\!(\phi_{\!B\!A})\mathbf{v}\Big\rvert^{2}\!}{ \|\mathbf{q}\|^2 \|\mathbf{q}-\mathbf{w}_l\|^2}, \;l \in \mathcal{N}_u,
\label{gamma}
\end{align}
where $\bar{P}=\frac{P}{\sigma^2}$, and 
\begin{equation}
G_{\text{BF}\!,l}(\mathbf{q},\boldsymbol{\psi}\!,\boldsymbol{\theta})\!=\!\Big\lvert
\mathfrak{\mathbf{a}}_2^{H}\!(\phi_{\!AU\!,\,l}^{(e)}, \phi_{\!AU\!,\,l}^{(a)})\mathbf{\Theta}
\mathfrak{\mathbf{a}}_{I}(\vartheta_{\!A\!B}^{(e)}\!,\vartheta_{\!A\!B}^{(a)})\Big\rvert^{2\!} \!\! =\! \Big\lvert\mathbf{f}_{l}^H\boldsymbol{\theta}\Big\rvert^{2}  
\end{equation}
represents the passive beamforming gain at user $l$, where $\boldsymbol{\theta}=[e^{j\theta_1},\cdots,e^{j\theta_N}]^T$ and
\begin{align}
\mathbf{f}_{l}=&\mathfrak{\mathbf{a}}_2(\phi_{\!AU\!,\,l}^{(e)}, \phi_{\!AU\!,\,l}^{(a)}) \odot \mathfrak{\mathbf{a}}_{I}^*(\vartheta_{\!A\!B}^{(e)},\vartheta_{\!A\!B}^{(a)}) \notag \\
=&(\mathfrak{\mathbf{a}}_{2x,l} \otimes \mathfrak{\mathbf{a}}_{2y,l}) \odot (\mathfrak{\mathbf{a}}_{Ix}^* \otimes \mathfrak{\mathbf{a}}_{Iy}^*) \notag \\
=&(\mathfrak{\mathbf{a}}_{2x,l} \odot \mathfrak{\mathbf{a}}_{Ix}^*) \otimes (\mathfrak{\mathbf{a}}_{2y,l} \odot \mathfrak{\mathbf{a}}_{Iy}^*) \notag \\
=&\mathbf{f}_{lx} \otimes \mathbf{f}_{ly}. \label{fl} 
\end{align}
Note that as all users share a common BS-IRS channel, to maximize the received SNR at any user $l$ in \eqref{gamma}, the optimal transmit beamforming vector is given by 
\begin{equation}	\label{v*}
\mathbf{v}^{\text{opt}} = \frac{\mathfrak{\mathbf{a}}_B(\phi_{BA})}{\|\mathfrak{\mathbf{a}}_B(\phi_{BA})\|}=\frac{\mathfrak{\mathbf{a}}_B(\phi_{BA})}{\sqrt{M}}.
\end{equation}
Substituting \eqref{v*} into \eqref{gamma} yields
\begin{equation}	
\gamma_l(\mathbf{q}, \boldsymbol{\psi},\boldsymbol{\theta}) =
\frac{\bar{P}\beta_0^2M F_{\text{AG},l}(\mathbf{q}, \boldsymbol{\psi}) 
G_{\text{BF},l}(\mathbf{q},\boldsymbol{\psi},\boldsymbol{\theta})}
{  \|\mathbf{q}\|^2 \|\mathbf{q}-\mathbf{w}_l\|^2}. \label{finalgamma0} 
\end{equation}

\subsection{Problem Formulation}

The goal of this paper is to maximize the minimum received SNR among all users by jointly optimizing the AIRS's location $\mathbf{q}$, 3D orientation angles $\boldsymbol{\psi}$, and phase shifts $\boldsymbol{\theta}$. 
Hence, the optimization problem can be formulated as
\begin{align}	\notag
(\mathrm{P}1)\; &\max_{\mathbf{q},\boldsymbol{\psi},\boldsymbol{\theta}}\;\;\min_{l \in \mathcal{N}_u}\;\;\gamma_l(\mathbf{q},\boldsymbol{\psi},\boldsymbol{\theta}) \\
\mathrm{s.t.} \;
\;&\mathbf{q} \in {\cal Q}, \\
\;&q_xL_1+q_yL_2+HL_3 \ge 0, \label{stp21} \\
\;(&q_x\!-\!w_{lx})L_1\!+\!(q_y\!-\!w_{ly})L_2\!+\!HL_3 \ge 0, \label{stp22} \\
\;&\big|(\boldsymbol{\theta})_n\big| = 1, 
\label{stp23}
\end{align}
where ${\cal Q}$ denotes a prescribed region for the AIRS's movement. The constraints \eqref{stp21} and \eqref{stp22} are imposed to ensure that the BS and all users are located in the half-reflection plane of the AIRS, i.e., \eqref{cosphi1} and \eqref{cosphi2} must be positive. 
However, problem (P1) is a non-convex optimization problem with the design variables $\mathbf{q}$, $\boldsymbol{\psi}$ and $\boldsymbol{\theta}$ intricately coupled with each other. To gain useful insights into the proposed passive 6DMA, we first consider a simplified single-user scenario in Section III and then address the general multi-user scenario in Section IV.

\section{Single-user Case}
In this section, we focus on solving (P1) under the single-user case with $N_u=1$. Without loss of generality, we label the single user as user 0 and assume that it is located along the $x$-axis. As such, let $\mathbf{w}_0=[D,0,0]^T$ denote its coordinate, where $D$ denotes its distance with the BS. Furthermore, to reduce the end-to-end path loss and simplify the optimization, we further set $q_y=0$ for the AIRS in this section.\footnote{Notably, $q_y$ can also be optimized by performing a similar algorithm as in Section IV. However, assuming a fixed $q_y$ helps reveal more essential insights into the effects of the AIRS orientation, as will be shown later in this section.}

\subsection{Optimal AIRS Reflection for Given Location and Orientation}
\begin{figure}[tb]
\centerline{\includegraphics[width=0.4350\textwidth]{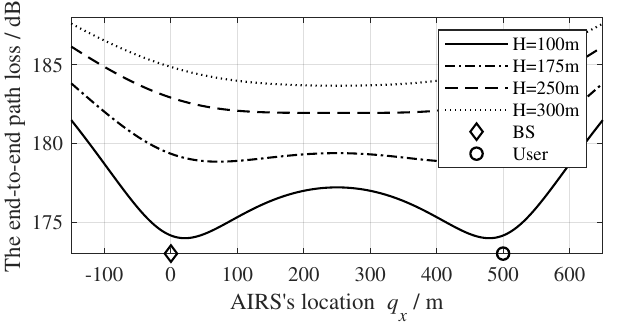}}
\caption{The end-to-end path loss versus AIRS's location with $D=500$ m.}
\label{pl}
\end{figure}

\begin{figure}[tb]
\centerline{\includegraphics[width=0.4533\textwidth]{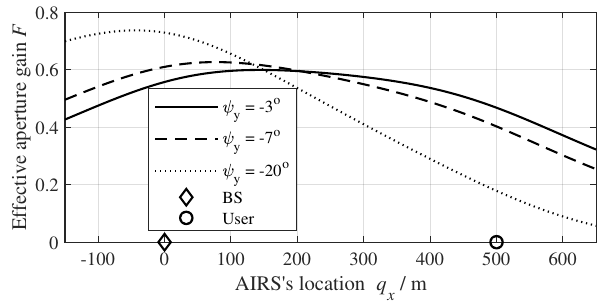}}
\caption{Effective aperture gain versus AIRS's location with $\psi_z = \psi_x = 0$ and $H=300$ m.}
\label{eal}
\end{figure}

In the case of a single user, the AIRS reflection should be designed to maximize the end-to-end channel power gain, i.e., $G_{\text{BF},0}(\mathbf{q},\boldsymbol{\psi},\boldsymbol{\theta})$. Thus, the optimal phase shift of the AIRS's $n$-th reflecting element is given by
\begin{equation}	\label{thetan}
\theta_n = \angle  \big( \mathfrak{\mathbf{a}}_2(\phi_{AU\!,\,0}^{(e)},\phi_{AU\!,\,0}^{(a)})\big)_n  -\angle\big( \mathfrak{\mathbf{a}}_I(\vartheta_{AB}^{(e)},\vartheta_{AB}^{(a)})\big)_n,
\end{equation}
such that the signals reflected by all reflecting elements of the AIRS are in-phase at the user's receiver. By noting $N_u=1$ and $q_y=0$, \eqref{F} reduces to 
\begin{equation}
F_{\text{AG},0}(q_x,\boldsymbol{\psi})=\frac{q_xL_1+HL_3}{\sqrt{q_x^2+H^2}}\frac{(q_x\!-\!D)L_1+HL_3}{\sqrt{(q_x\!-\!D)^2+H^2}}. \label{Fsingle}
\end{equation}
By substituting \eqref{thetan} and \eqref{Fsingle} into \eqref{finalgamma0}, we have
\begin{equation}	\label{finalgamma}
\gamma_0(q_x,  \boldsymbol{\psi}) = \frac{\bar{P}\beta_0^2F_{\text{AG},0}(q_x,\boldsymbol{\psi}) M N^2}{ [q_x^2 +H^2][(q_x-D)^2+H^2]}.
\end{equation}
It is observed from \eqref{finalgamma} that the maximum received SNR depends on the end-to-end path loss as well as the effective aperture gain $F_{\text{AG},0}(q_x,\boldsymbol{\psi})$, both of which are affected by the AIRS's location, $q_x$. To better illustrate the effect of $q_x$ on them, we plot in Figs.~\ref{pl} and \ref{eal} the end-to-end path loss and effective aperture gain versus $q_x$, respectively, with $\beta_0=-40$ dB.

It is observed from Fig.~\ref{pl} that there exists an optimal $q_x$ such that the end-to-end path loss is minimized, and such an optimal location depends on the UAV's altitude, $H$. As $H$ is low/high, the optimal $q_x$ will approach the end points/midpoint of the BS-user line, as also previously revealed in \cite{AIRS}. On the other hand, it is observed from Fig.~\ref{eal} that there also exists an optimal $q_x$ maximizing the effective aperture gain, which is affected by the orientation angles, e.g., $\psi_y$. In particular, when $\psi_y=-20^\circ$, the optimal $q_x$ may be even at the left hand side (LHS) of the BS, instead of the end points or midpoint of the BS-user segment. It is seen that the optimal $q_x$ for minimizing the end-to-end path loss and maximizing the effective aperture gain may be different. In addition, the IRS's orientation angle will also affect the optimal $q_x$ for maximizing the effective aperture gain. Thus, the IRS's location and orientation should be jointly optimized to balance the end-to-end path loss and effective aperture gain.

Given the user's received SNR in \eqref{finalgamma}, problem (P1) can be simplified as
\begin{align}	\notag
(\mathrm{P}2)\; \max_{q_x,\boldsymbol{\psi}}\;\;&\gamma_0(q_x,\boldsymbol{\psi}) \\
\mathrm{s.t.} \;\;&q_xL_1+HL_3 \ge 0, \notag \\
\; (&q_x-D)L_1+HL_3 \ge 0, \notag \\
\; &q_x \in {\cal Q}. 
\notag
\end{align}
However, problem ($\mathrm{P}2$) is still difficult to be optimally solved due to the coupling between the location $q_x$ and orientation $\boldsymbol{\psi}$ in the objective function. Although an optimal solution can be obtained by performing an exhaustive search over $q_x$ and $\boldsymbol{\psi}$, this incurs practically exorbitant complexity. Next, we will propose an efficient algorithm to solve ($\mathrm{P}2$) optimally.

\subsection{Proposed Solutions to (P2)}
In this subsection, we first show that for any given AIRS's location, its 1D orientation around the $y'$-axis  suffices to achieve the optimal performance of 3D orientation, and then solve the resulting simplified optimization problem accordingly.

\subsubsection{Optimality of AIRS's 1D Orientation}
To show the optimality of 1D orientation, we present the following proposition to first show the optimality of two-dimensional (2D) orientation.

\newtheorem{proposition}{Proposition}
\begin{proposition}  
For any given AIRS's 3D orientation $\boldsymbol{\psi}_{\text{3D}} = [\psi_z, \psi_y, \psi_x]^T$ and location $q_x$, there always exists a 2D orientation solution $\boldsymbol{\psi}_{\text{2D}}^\star = [0,\psi_y^\star,\psi_x^\star]^T$, such that
\begin{equation} \label{prop1}
\gamma_0(q_x,\boldsymbol{\psi}_{\text{2D}}^\star) = \gamma_0(q_x,\boldsymbol{\psi}_{\text{3D}}).
\end{equation}
\end{proposition}

The detailed proof is provided in Appendix.
Following the result in Proposition 1, we further show the optimality of 1D orientation in Proposition 2 below.
\begin{proposition} \label{p2}
For any given 2D orientation solution $\boldsymbol{\psi}_{\text{2D}}=[0,\psi_y,\psi_x]^T$ and $q_x$, there always exists a 1D orientation solution $\boldsymbol{\psi}_{\text{1D}}^\star=[0, \psi_y^\star, 0]^T$ that satisfies
\begin{equation} \label{prop2}
\gamma_0(q_x, \boldsymbol{\psi}_{\text{1D}}^\star) \ge \gamma_0(q_x,\boldsymbol{\psi}_{\text{2D}}).
\end{equation}
\end{proposition}
\begin{IEEEproof}
Similar to the proof of Proposition 1, it is equivalent to prove $F_{\text{AG},0}(q_x,\boldsymbol{\psi}_{\text{1D}}^\star) \ge  F_{\text{AG},0}(q_x,\boldsymbol{\psi}_{\text{2D}})$, for which it suffices to prove $\cos \phi_1(q_x,\boldsymbol{\psi}_{\text{1D}}^\star) \ge \cos \phi_1(q_x,\boldsymbol{\psi}_{\text{2D}})$ and $\cos \phi_{2,0}(q_x,\boldsymbol{\psi}_{\text{1D}}^\star) \ge \cos \phi_{2,0}(q_x,\boldsymbol{\psi}_{\text{2D}})$. Note that 
\begin{align} \label{lef32}
\cos\!\phi_1(q_x,\boldsymbol{\psi}_{\text{1D}}^\star) =& \;\frac{q_x\sin\psi_y^\star+H\cos\psi_y^\star}{\sqrt{q_x^2+H^2}} \notag \\
 \overset{\text{(a)}}{\ge}& \;\frac{(q_x\sin\psi_y^\star +H\cos\psi_y^\star)\cos\psi_x^\star}{\sqrt{q_x^2+H^2}} \notag \\
=&\;\cos\phi_1(q_x,\boldsymbol{\psi}_{\text{2D}}),\\
\cos\!\phi_{2,0}(q_x,\boldsymbol{\psi}_{\text{1D}}^\star) =& \;\frac{(q_x-D)\sin\psi_y^\star+H\cos\psi_y^\star}{\sqrt{(q_x-D)^2+H^2}}  \notag \\
\overset{\text{(b)}}{\ge}&\;\frac{\big((q_x-D)\sin\psi_y^\star+H\cos\psi_y^\star\big)\cos\psi_x^\star}{\sqrt{(q_x-D)^2+H^2}}\notag \\
=&\; \cos\phi_{2,0}(q_x,\boldsymbol{\psi}_{\text{2D}}),  \label{lef32a}
\end{align}
where inequalities (a) and (b) hold in that $q_x\sin\psi_y^\star+H\cos\psi_y^\star\ge 0$ and $(q_x-D)\sin\psi_y^\star+H\cos\psi_y^\star\ge 0$ due to the constraints in (P2) for effective signal reflection.
Thus, the proof of Proposition 2 is completed.
\end{IEEEproof}

Proposition 2 suggests that only the 1D orientation around the $y'$-axis, i.e., $\psi_y$, is sufficient to achieve the maximum received SNR at the user. In this case, the effective aperture gain in \eqref{Fsingle} can be simplified as 
\begin{equation} 
F_{\text{AG},0}(q_x,\psi_y)\!=\!\frac{q_x\!\sin\psi_y\!+\!H\!\cos\psi_y}{\sqrt{q_x^2+H^2}} \frac{(q_x\!-\!D)\!\sin\psi_y\!+\!H\!\cos\psi_y}{\sqrt{(q_x-D)^2+H^2}}, \label{Fpsi+} 
\end{equation}
and the user's received SNR in \eqref{finalgamma} becomes 
\begin{equation}	
\gamma_{b}(q_x,\psi_{y}) \!=\! \frac{\!\bar{P}\beta_{0}^{2\!} M\!N^{2}\!\sin\!\big(\psi_y\!+\!\psi_1(q_x)\!\big)\!\sin\!\big(\psi_y\!+\!\psi_2(q_x)\!\big)\!}{ [q_x^2 +H^2][(q_x-D)^2+H^2]},\label{gamma1D}
\end{equation}
where 
\begin{align} \label{siny1}
\psi_1(q_x) &= \arccos\frac{q_x}{\sqrt{q_x^2+H^2}}, \\
\psi_2(q_x) &= \arccos\frac{q_x-D}{\sqrt{(q_x-D)^2+H^2}}.  \label{siny2} 
\end{align}
The formulated problem ($\mathrm{P}2$) can be simplified as
\begin{align} 
(\mathrm{P3})\max_{q_x,\psi_y} \;\;&\gamma_{b}(q_x,\psi_y) \notag \\
\mathrm{s.t.}\;\; &q_x\sin\psi_y + H\cos\psi_y \ge 0, \notag \\
(&q_x-D)\sin\psi_y + H\cos\psi_y \ge0, \notag \\
 &q_x \in {\cal Q}. \notag 
\end{align}
Based on \eqref{Fpsi+} and \eqref{gamma1D}, it is noted that without the AIRS orientation optimization, i.e., $\psi_y\!=\!0$, we have $\gamma_{b}(q_x,0)\!=\! \frac{\bar{P}\beta_0^2 M N^2 H^2}{ [q_x^2 +H^2]^{3/2}[(q_x-D)^2+H^2]^{3/2}}$.\vspace{1pt}  In this case, the optimal $q_x$ should be such that the end-to-end path loss is minimized, which is identical to the isotropic signal reflection as studied in \cite{AIRS}, i.e.,
\begin{equation} \label{oqx}
q_x^{\text{iso}} = \begin{cases}
\frac{D}{2},  & \text{ if } 0 \le \frac{D}{H} \le 2 \\
\frac{D}{2}\pm\sqrt{\frac{D^2}{4}-H^2},  & \text{ otherwise.} 
\end{cases}
\end{equation} 

\subsubsection{AIRS's Optimal 1D Orientation for Any Given Location} \label{IIIB}
For the simplified problem (P3), we have the following proposition that characterizes the optimal 1D orientation in terms of the AIRS's location, $q_x$.

\begin{proposition}
For any given $q_x$ in (P3), the optimal 1D orientation for (P3) is given by
$\psi_y^\star(q_x) = \frac{\pi - \psi_1(q_x)- \psi_2(q_x)}{2}$.
\end{proposition}
\begin{IEEEproof} 
By exploiting the product-to-sum identities, \eqref{gamma1D} becomes 
\begin{align} \label{gamma1Da}
\gamma_{b}(q_x, \psi_y) \!=\! \alpha_{0}(q_x) \Big[&\cos\big(\psi_1(q_x) - \psi_2(q_x)\big)-\notag \\
&\cos\big(2\psi_y + \psi_1(q_x)+\psi_2(q_x)\big)\Big], 
\end{align}
where $\alpha_{0}(q_x)\!=\!\frac{\bar{P}\beta_0^2 M N^2}{2[q_x^2 +H^2][(q_x-D)^2+H^2]}$. To maximize \eqref{gamma1Da} for any given $q_x$, the optimal $\psi_y$ should satisfy $\cos\big(2\psi_y^\star+ \psi_1(q_x)+\psi_2(q_x)\big)=-1$, which results in
\begin{equation} \label{psiy*}
\psi_y^\star(q_x) = \frac{\pi - \psi_1(q_x) - \psi_2(q_x)}{2}.
\end{equation}
This completes the proof.
\end{IEEEproof}

The analytical optimal solution for 1D orientation in \eqref{psiy*} reveals several interesting insights.  Specifically, if the AIRS is located above the midpoint of the BS and the user, i.e., $q_x= \frac{D}{2}$, we have $\psi_1(q_x) +\psi_2(q_x) = \pi $ by recalling \eqref{siny1} and \eqref{siny2}, which results in $\psi_y^\star(q_x)=0$. This implies that no orientation is needed in this case.
Interestingly, if the AIRS is sufficiently far from the BS/user, i.e., $q_x \to \infty$ (or $-\infty$), we have $\psi_1(q_x)=\psi_2(q_x)=0$ (or $\pi$), which results in $\psi_y^\star(q_x) = \frac{\pi}{2}$ (or $-\frac{\pi}{2}$). It is also worth noting that $\psi^\star_y\big(\frac{D}{2}\!+\delta_q\big)=\psi^\star_y\big(\frac{D}{2}\!-\delta_q\big)$ for any $\delta_q >0$, which implies that the optimal AIRS's orientation angle is symmetric to $q_x=\frac{D}{2}$. 

\subsubsection{AIRS's Optimal Location}
By substituting the optimal $\psi_y$ in \eqref{psiy*} into (P3), we obtain the following single scalar-variable optimization problem w.r.t. $q_x$, i.e., 
\begin{align} 
(\mathrm{P4})\;\max_{q_x}\;&\;\gamma_{c}(q_x) \notag \\
\mathrm{s.t.}&\; q_x \in {\cal Q}, \notag 
\end{align}
where
\begin{equation} \label{gamma1Db}
\gamma_{c}(q_x) = \frac{\bar{P}\beta_0^2 M N^2\Big[1\!+\!\cos\big(\psi_1(q_x)\!-\!\psi_2(q_x)\big)\Big]}{2\, [q_x^2 +H^2][(q_x-D)^2+H^2]}.
\end{equation}
It is noted from \eqref{gamma1Db} that the AIRS's orientation mainly affects the user's received SNR through the term $1\!+\!\cos\big(\psi_1(q_x)\!-\!\psi_2(q_x)\big)$. To gain more insights, we investigate the following special cases.

First, in the case of a terrestrial IRS with $H\! \rightarrow \!0$, we have $\psi_1(q_x)=\psi_2(q_x)=0$ based on \eqref{siny1} and \eqref{siny2}, and thus $1+\cos\big(\psi_1(q_x)-\psi_2(q_x)\big)$ can obtain its maximum value of $2$. As a result,  the optimal $q_x$ should be such that the end-to-end path loss, i.e., the denominator of \eqref{gamma1Db}, is minimized, which is given by $q^\star_x=0$ or $D$. In this case, the optimal orientation angle for the terrestrial IRS is always $\psi_y^\star(q_x)=\pm\frac{\pi}{2}$ based on \eqref{psiy*}, which is consistent with the result presented in \cite{ORI1}. 

On the other hand, if the AIRS's altitude is sufficiently high, i.e., $H \!\rightarrow\! \infty$, we have $\psi_1(q_x)\!=\!\psi_2(q_x)\!=\!\frac{\pi}{2}$, and the optimal $q_x$ is given by $q_x^\star\!=\!0$ or $D$, similar to the case of a terrestrial IRS. However, in this case, the AIRS's optimal orientation angle is given by $\psi_y^\star(q_x^\star)\!=\!0$. This implies that optimizing the AIRS's orientation can barely bring performance gain, which is expected as the AIRS may be treated as a point in this case due to its extremely far distance with the ground. 

Finally, if the BS and the user are sufficiently close, i.e., $D\!\rightarrow\!0$, we have $\psi_1(q_x)\!=\!\psi_2(q_x)$ and $1+\cos\big(\psi_1(q_x)-\psi_2(q_x)\big)\!=\!2$. As a result, the AIRS's optimal location should minimize the end-to-end path loss, which is given by $q_x^\star=0$. Note that this also results in $\psi_1(q_x)\!=\!\psi_2(q_x)\!=\!\frac{\pi}{2}$, and $\psi_y^\star(q_x^\star)\!=\!0$, implying that $D\! \rightarrow \!0$ shows similar trend to $H\! \rightarrow \!\infty$.

However, in other general cases with an arbitrary $H$ or $D$, \eqref{gamma1Db} is a highly non-convex function w.r.t. $q_x$, making it  difficult to obtain its optimal solution in closed form. Hence, we apply an exhaustive search over ${\cal Q}$ to obtain the resulting $q_x$.

\section{General Multi-user Case}
In this section, we aim to solve (P1) in the general multi-user case with $N_u > 1$. To this end, we propose an AO algorithm to decouple it into three sub-problems w.r.t. the AIRS's location $\mathbf{q}$, orientation $\boldsymbol{\psi}$, and phase shifts $\boldsymbol{\theta}$, respectively, as presented in Section IV-A. To avoid the local convergence issue of the AO algorithm, an enhanced AO algorithm is also proposed by applying the GS in Section IV-B. 

\subsection{AO Algorithm without GS}
Consider the $j$th iteration of the AO algorithm and denote the initial values of the AIRS's location, orientation, and phase shifts as $\mathbf{q}^{(j-1)}$, $\boldsymbol{\psi}^{(j-1)}$ and $\boldsymbol{\theta}^{(j-1)}$, respectively.

\subsubsection{Location Optimization with Given AIRS Orientation and Phase Shifts}
In this case, (P1) is simplified as
\begin{align}
\text{(P1.1)}\; &\max_{\mathbf{q}}\; \min_{l \in \mathcal{N}_u} \;\gamma_l(\mathbf{q},\boldsymbol{\psi}^{(j-1)},\boldsymbol{\theta}^{(j-1)}) \notag \\
\text{s.t.}\;&\; \mathbf{q} \in {\cal Q}. 
\end{align}
Problem (P1.1) is still a non-convex optimization problem w.r.t. the AIRS's location $\mathbf{q}$. One straightforward approach is by discretizing the space $\mathcal{Q}$ and searching for the optimal location of the AIRS among the sampling points. However, this results in practically high computational complexity, especially in the case of a high sampling resolution to improve the searching accuracy. To properly balance the complexity and performance, we propose a 2D hybrid coarse- and fine-grained search strategy to optimize $\mathbf{q}$, as depicted in Fig.~\ref{locori}(a). 

Specifically, $\mathcal{Q}$ is first uniformly divided into $\bar{N}_x \times \bar{N}_y$ rectangular sub-regions along the $x$- and $y$-axes, respectively, where $\bar{N}_x$ and $\bar{N}_y$ denote the numbers of the sampling points along $x$- and $y$-axes, respectively. As such, there exist $N_{\text{tot}}=\bar{N}_x \times \bar{N}_y$ sub-regions in total. Denote by $\mathbf{q}_n \in \mathcal{Q}$ the center of the $n$-th sub-region, with $1 \le n \le N_{\text{tot}}$. Then, the minimum received SNR among all users if the AIRS is deployed above $\mathbf{q}_n$ can be expressed as
\begin{equation}
\gamma_{\min}(\mathbf{q}_n, \boldsymbol{\psi}^{(j-1)},\boldsymbol{\theta}^{(j-1)}) = \min_{l \in \mathcal{N}_u} \; \gamma_l(\mathbf{q}_n,\boldsymbol{\psi}^{(j-1)},\boldsymbol{\theta}^{(j-1)}). \label{WCSNR1}
\end{equation}
Among the $N_{\text{tot}}$ sub-regions, we denote $\mathbf{q}_{n^*}$ as the best center that achieves the highest minimum received SNR among all centers, with
\begin{equation}
n^*=\arg\max_{1 \le n \le N_{\text{tot}}} \gamma_{\min}(\mathbf{q}_n, \boldsymbol{\psi}^{(j-1)},\boldsymbol{\theta}^{(j-1)}). \label{n*}
\end{equation}
Note that the sampling resolution in the above searching can be set relatively low to quickly determine the sub-region in \eqref{n*}. Next, a finer-grained search within the $n^*$-th sub-region is conducted by discretizing it into a multitude of sampling points, as shown in Fig.~\ref{locori}(a). Let $N_{x,n^*}$ and $N_{y,n^*}$ denote the number of sampling points along $x$- and $y$-axes within the $n^*$-th sub-region. As such, there are $N_{\text{tot}, n^*}=N_{x,n^*} \times N_{y, n^*}$ sampling points in the $n^*$-th sub-region, and we denote by $\mathbf{q}_{n^*}^{(m)} \in \mathcal{Q}$ the coordinate of the $m$-th sampling point in it. Then, the optimized AIRS location in the $j$-th AO iteration can be obtained as 
\begin{equation}
\mathbf{q}^{(j)} \!=\! \mathbf{q}_{n^*}^{(m^{\star})}\!\!,\; m^{\star} \!=\! \arg \max_{1 \le m \le N_{\text{tot},n^*}} \gamma_{\min}(\mathbf{q}_{n^*}^{(m)}\!,\boldsymbol{\psi}^{(j\!-\!1)}\!,\boldsymbol{\theta}^{(j\!-\!1)}). \label{qj}
\end{equation}

\subsubsection{Orientation Optimization with Given AIRS Location and Phase Shifts} 

Next, the sub-problem of AIRS orientation optimization can be formulated as
\begin{align}
\text{(P1.2)}\; &\max_{\boldsymbol{\psi}} \;\min_{l \in \mathcal{N}_u} \; \gamma_l(\mathbf{q}^{(j)},\boldsymbol{\psi},\boldsymbol{\theta}^{(j-1)}) \notag \\
\text{s.t.}\;\;& q_{x}^{(j)}L_1+q_{y}^{(j)}L_2+HL_3 \ge 0, \label{st321}\\
\;(&q_{x}^{(j)}\!-\!w_{lx})L_1\!+\!(q_{y}^{(j)}\!-\!w_{ly})L_2\!+\!HL_3 \ge 0, \label{st322} 
\end{align}
where $[q_x^{(j)},q_y^{(j)},H]^T = \mathbf{q}^{(j)}$. Unlike the single-user case, problem (P1.2) is difficult to be optimally solved, since the optimal AIRS needs to cater to the SNRs at multiple users at the same time. To tackle the issue, we apply a similar hybrid coarse- and fine-grained searching method as in AIRS position optimization. 

\begin{figure}[tb]
\centerline{
\includegraphics[width=0.4702\textwidth]{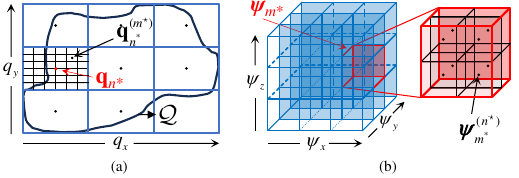}}
\caption{Schematic diagram of the hybrid grained search.}
\label{locori}
\end{figure} 

Specifically, we uniformly partitioned the angular interval $[-\pi/2, \pi/2]$ around $z'$-, $y'$-, $x'$-axes into $\tilde N_z$, $\tilde N_y$, and $\tilde N_x$ segments, respectively, which give rise to $\tilde N_{\text{tot}} =\tilde N_z \times \tilde N_y \times \tilde N_x$ cuboid in total, as shown in Fig.~\ref{locori}(b). Denote by $\boldsymbol{\psi}_m = [\tilde \psi_z, \tilde \psi_y, \tilde \psi_x]^T$ the center of the $m$-th cuboid, with $ 1 \le m \le \tilde N_{\text{tot}}$, where 
\begin{align}
\tilde \psi_z = \frac{\pi(2n_z+1-\tilde N_z)}{2\tilde N_z}, &\; n_z = \lfloor m / (\tilde N_y \tilde N_x)\rfloor, \notag \\
\tilde \psi_y = \frac{\pi(2n_y+1-\tilde N_y)}{2\tilde N_y}, &\; n_y = \lfloor \big(m-n_z(\tilde N_y\tilde N_x)\big) / \tilde N_x \rfloor, \notag \\
\tilde \psi_x = \frac{\pi(2n_x-1-\tilde N_x)}{2\tilde N_x}, &\; n_x = m-n_z(\tilde N_y\tilde N_x)-n_y\tilde N_x, \notag 
\end{align} 
where $\lfloor \cdot \rfloor$ denotes the largest integer that is no larger than its argument. Among them, we denote $\boldsymbol{\psi}_{m^*}$ as the best cuboid center that achieves the highest minimum received SNR among all centers, with
\begin{equation}
m^* = \arg \max_{1 \le m \le \tilde N_{\text{tot}}} \gamma_{\min}(\mathbf{q}^{(j)}, \boldsymbol{\psi}_m, \boldsymbol{\theta}^{(j-1)}).
\end{equation}
Next, a finer-grained search within the $m^*$-th cuboid  is conducted by further discretizing it into a multitude of sampling points, as shown in Fig.~\ref{locori}(b). Let $\tilde N_{z,m^*}$, $\tilde N_{y,m^*}$, and $\tilde N_{x,m^*}$ denote the number of sampling angles of $\psi_z$, $\psi_y$, and $\psi_z$ within the $m^*$-th cuboid, respectively. As such, there are $\tilde N_{\text{tot},m^*}=\tilde N_{z,m^*} \times \tilde N_{y,m^*} \times\tilde N_{x,m^*}$ sampling points in the $m^*$-th cuboid, and we denote by $\boldsymbol{\psi}_{m^*}^{(n)}$ the coordinate of the $n$-th sampling point in it. Then, the optimized AIRS orientation in the $j$-th AO iteration can be obtained as
\begin{equation}
\boldsymbol{\psi}^{(j)} \!= \boldsymbol{\psi}_{m^*}^{(n^\star)}\!,\;
n^{\star} = \arg \max_{1 \le n \le \tilde N_{\text{tot}, m^*}} \gamma_{\min}(\mathbf{q}^{(j)}, \boldsymbol{\psi}_{m^*}^{(n)}, \boldsymbol{\theta}^{(j\!-\!1)}). \label{psij}
\end{equation}

\subsubsection{Phase Shift Optimization with Given AIRS Location and Orientation}

Finally, the sub-problem of AIRS phase-shift optimization is given by
\begin{align}
\text{(P1.3)}\;& \max_{\boldsymbol{\theta}}\; \min_{l \in \mathcal{N}_u} \;\gamma_l(\mathbf{q}^{(j)},  \boldsymbol{\psi}^{(j)}, \boldsymbol{\theta}) \notag \\
\text{s.t.} \;\; & \big\lvert (\boldsymbol{\theta})_n \big\rvert = 1, \; n=1,2,\cdots, N.
\end{align}

Problem (P1.3) is similar to conventional IRS-aided multicast as studied in e.g., \cite{SDR} and \cite{Theta1}. However, as the AIRS is generally equipped with a larger number of reflecting elements than its terrestrial counterparts, the existing methods, e.g., semi-definite relaxation (SDR), may result in excessively high computational complexity. To address this difficulty, we further decouple problem (P1.3) into two sub-problems corresponding to the horizontal and vertical passive beamforming of the AIRS along the $x'$- and $y'$-axes, respectively, and solve them alternately. Let $\boldsymbol{\theta}_x$ and $\boldsymbol{\theta}_y$ denote the horizontal and vertical AIRS passive beamforming, respectively, with $\boldsymbol{\theta}\!=\!\boldsymbol{\theta}_x \otimes \boldsymbol{\theta}_y$. By noting $\mathbf{f}_{lx}$ and $\mathbf{f}_{ly}$ in \eqref{fl}, the received SNR in \eqref{finalgamma0} can be re-expressed in terms of $\boldsymbol{\theta}_x$ and $\boldsymbol{\theta}_y$ as
\begin{equation}
\gamma_l(\mathbf{q}^{(j)}\!\!,  \boldsymbol{\psi}^{(j)}\!\!, \boldsymbol{\theta})=\frac{\bar{P}\beta_0^2M F_{\text{AG},l}(\mathbf{q}^{(j)}\!\!, \boldsymbol{\psi}^{(j)})
\lvert \mathbf{f}_{lx}^H\boldsymbol{\theta}_x\rvert^2 \lvert \mathbf{f}_{ly}^H\boldsymbol{\theta}_y \rvert^2}
{  \|\mathbf{q}^{(j)}\|^2 \|\mathbf{q}^{(j)}-\mathbf{w}_l\|^2}. 
\end{equation}

For any given vertical passive beamforming $\boldsymbol{\theta}_y$ and by discarding irrelevant constant terms, it can be shown that the horizontal passive beamforming design can be formulated as
\begin{align}
(\text{P1.3$x$})&\;\max_{\boldsymbol{\theta}_x}\;\min_{l \in \mathcal{N}_u}\; \alpha_l \lvert \mathbf{f}_{lx}^H\boldsymbol{\theta}_x \rvert^2 \notag \\
\text{s.t.}&\;\; \big| (\boldsymbol{\theta}_x)_n  \big| = 1, \; n=1,2,\cdots, N_x, 
\end{align}
where 
\begin{equation}
\alpha_l = \sqrt{\cfrac{\cfrac{F_{\text{AG},l}(\mathbf{q}^{(j)}, \boldsymbol{\psi}^{(j)}) }{ \|\mathbf{q}^{(j)}-\mathbf{w}_l\|^2}}{ \sum_{l=1}^{N_u} \cfrac{F_{\text{AG},l}(\mathbf{q}^{(j)},\boldsymbol{\psi}^{(j)}) }{ \|\mathbf{q}^{(j)}-\mathbf{w}_l\|^2
}}},\; l\in \mathcal{N}_u, \label{alphal}
\end{equation}
is a normalized constant expression depending on $\mathbf{q}^{(j)}$ and $\boldsymbol{\psi}^{(j)}$. By introducing an epigraph auxiliary variable $\delta$, (P1.3$x$) can be recast as
\begin{align}
\max_{\boldsymbol{\theta_x}}\;\; &\delta \label{tarfun} \\
\text{s.t.} \;\; & \alpha_l \big | \mathbf{f}_{lx}^H \boldsymbol{\theta}_x \big |^2 \ge \delta, \; l \in \mathcal{N}_u,  \label{cvxst1} \\
& \big| (\boldsymbol{\theta}_x)_n \big| = 1,\; n=1,2,\cdots, N_x.		\label{cvxst2} 
\end{align}
However, \eqref{cvxst1} and \eqref{cvxst2} are both non-convex constraints. 
To recast \eqref{cvxst2} into a convex form, we first lift the beamforming vector $\boldsymbol{\theta}_x$ to a positive semi-defined (PSD) matrix $\mathbf{W} \in \mathbb{C}^{N_x \times N_x}$, with $\mathbf{W} = \boldsymbol{\theta}_x\boldsymbol{\theta}_x^H$ being a rank-one matrix. Due to the matrix lifting, we recast \eqref{cvxst1} into a convex form using auto-correlation function as \cite{Theta1}
\begin{equation}
\big | \mathbf{f}_{lx}^H \boldsymbol{\theta}_x \big |^2 = \text{Re}(\mathbf{f}_{lx}^H \mathbf{r}_x), \label{absre}
\end{equation}
where the auto-correlation vector is defined as
\begin{align}
(\mathbf{r}_x)_1 &= \sum_{k=1}^{N_x} (\boldsymbol{\theta}_x)_k (\boldsymbol{\theta}_x^*)_k = N_x, \label{rx1}\\
(\mathbf{r}_x)_n &= 2 \sum_{k=1}^{N_x-n+1} \mathbf{W}[k\!+\!n\!-\!1,k],\; n=2,3,\cdots,N_x. \label{rx2} 
\end{align}
Taking \eqref{absre}-\eqref{rx2} into account, \eqref{cvxst1} and \eqref{cvxst2} can be equivalently recast as
\begin{align}
\alpha_l \text{Re}(\mathbf{f}_{lx}^H \mathbf{r}_x)&\ge \delta,\;\, l \in \mathcal{N}_u, \label{st77}\\
\big| \mathbf{W}[n,n]\big| &= 1, \;n=1,2,\cdots,N_x, \label{st781}\\
\text{rank}(\mathbf{W}) &= 1. 	\label{st782} 
\end{align}
However, \eqref{st782} is still non-convex. To handle this issue, a penalty strategy is used to incorporate \eqref{st782} into the objective function \eqref{tarfun}. To be specific, for a PSD matrix $\mathbf{W}$ with $\text{Tr}(\mathbf{W})>0$, \eqref{st782} is equivalent to 
\begin{equation}
\|\mathbf{W}\|_*-\|\mathbf{W}\|_2 = 0,	\label{pnt}
\end{equation}
where $\|\mathbf{W}\|_*$ represents the nuclear norm, and $\|\mathbf{W}\|_2$ is the spectral norm. Taking \eqref{pnt} as the penalty term for rank-one matrix, \eqref{tarfun} can be expressed as
\begin{equation}
\max_{\mathbf{W}}\;\; \delta-\rho(\|\mathbf{W}\|_*-\|\mathbf{W}\|_2) \label{dcf},
\end{equation}
where $\rho>0$ is a pre-defined penalty parameter. Although \eqref{dcf} is still non-convex, the successive convex approximation (SCA) technique can be utilized to successively approach \eqref{dcf} with the first-order Taylor expansion at different local points. Specifically, in the $k$th SCA iteration, problem (P1.3$x$) can be expressed as
\begin{align}
\max_{\mathbf{W}} &\;\;\delta-\rho\bigg(\|\mathbf{W}\|_*-\bigg(\|\mathbf{W}^k\|_2\notag \\
&\;\;\;\;+ \text{Re}\Big(\text{Tr}\big((\partial_{\mathbf{W}^k}\|\mathbf{W}\|_2)(\mathbf{W}-\mathbf{W}^k)\big)\Big)\bigg)\bigg) \label{e85} \\
\text{s.t.}& \;\; \eqref{rx1},\; \eqref{rx2},\; \eqref{st77},\; \eqref{st781}, 
\notag
\end{align}
where $\mathbf{W}^k$ represents the local value of $\mathbf{W}$ in the $k$th SCA iteration, and the sub-gradient can be efficiently computed as $\partial_{\mathbf{W}^k}\|\mathbf{W}\|_2 = \mathbf{s}\mathbf{s}^H$, with $\mathbf{s}$ denoting the singular vector corresponding to the largest singular value of $\mathbf{W}^k$. The SCA proceeds until the objective values output by two adjacent iterations is smaller than a pre-defined threshold, or the number of iterations reaches a pre-defined value. Note that $\mathbf{W}^k$ is initialized as an all-zero matrix in this paper.

Upon the convergence of the SCA algorithm, the optimized horizontal AIRS passive beamforming can be obtained by performing the singular value decomposition (SVD) on the converged value of $\mathbf{W}$ as $\boldsymbol{\theta}_x^{(j)}$. 
Similarly to the above SCA procedures, we can obtain the SCA-optimized AIRS vertical passive beamforming as $\boldsymbol{\theta}_y^{(j)}$, for which the details are omitted for brevity. Finally, the optimized AIRS passive beamforming in the $j$-th AO iteration can be obtained as
\begin{equation}
\boldsymbol{\theta}^{(j)}=\boldsymbol{\theta}_x^{(j)} \otimes \boldsymbol{\theta}_y^{(j)}. \label{thetaj}
\end{equation}
The AO algorithm then proceeds to the $(j+1)$-th iteration. Since this gives rise to  non-decreasing objective value of (P1), the convergence can always be achieved.\cite{GS[23]}

\subsection{Enhanced AO with GS}
\begin{figure}[tb]
\centerline{
\includegraphics[width=0.4702\textwidth]{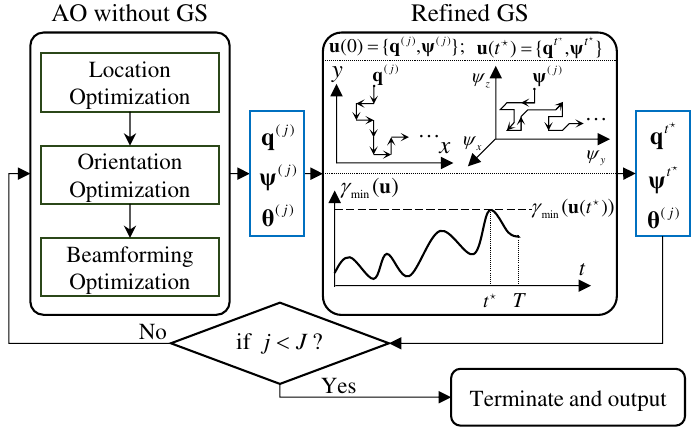}}
\caption{Schematic diagram of the proposed enhanced AO algorithm with GS.}
\label{algorithm}
\end{figure} 

Although the AO algorithm proposed in the last subsection is generally  effective to solve (P1), its ultimate performance may be locally optimal and even get trapped at an undesirable sub-optimal solution, especially for a max-min optimization problem\cite{QF}. To tackle the challenge, we propose an enhanced AO algorithm with GS phases in optimizing the AIRS's location and orientation\cite{GS[32],GS}, as depicted in Fig.~\ref{algorithm}.
Its basic idea is to explore nearby solutions around the solution obtained by solving \eqref{qj} and \eqref{psij} or jump to farther solutions with significantly different location and/or orientation via random selection. 
Note that the operations involved in the GS are conducted via a probability-based Markov chain.  This offers two major benefits: exploring nearby solutions enhances the stability of the max-min received SNR, while the random selection mitigates the risk of being stuck by low-quality local optima.

Mathematically, consider the end of the $j$-th AO iteration with $\boldsymbol{\theta}=\boldsymbol{\theta}^{(j)}$. Let $\mathbf{u} = \{\mathbf{q},\boldsymbol{\psi}\}$ denote the set of candidate location and orientation solutions with $\mathbf{u} \in \mathcal{S}$, where $\mathcal{S}$ denotes the set of all candidate location and orientation solutions in the GS. To generate $\mathcal{S}$, we equally partition the feasible space of the associated optimization variables (i.e., $q_x$, $q_y$, $\psi_z$, $\psi_y$, and $\psi_x$) into several sub-regions. The users' minimum received SNR at $\mathbf{u}$ can be expressed as
\begin{equation}
\gamma_{\min}(\mathbf{u}) =
 \min_{l \in \mathcal{N}_u} \; \gamma_l(\mathbf{u}, \boldsymbol{\theta}^{(j)} ).\label{eta79}
\end{equation}
Let $\Delta_q$ and $\Delta_\psi$ denote the spacing between any two adjacent location variables (i.e., $q_x$ and $q_y$) and that between two adjacent orientation variables (i.e., $\psi_z$, $\psi_y$, and $\psi_x$) after the space partitioning, respectively.

\begin{algorithm}[tb]
\caption{Proposed Enhanced AO with GS for Solving (P1).}\label{alg:alg1}
\begin{algorithmic}
\STATE Input: $N_x$, $N_y$, $J$, $T$, $N_u$, $\mathbf{w}_l$
\STATE $j \leftarrow 1$
\STATE \textbf{while} $j<J$:
\STATE \hspace{0.5cm} // AO without GS
\STATE \hspace{0.5cm} Update $\mathbf{q}^{(j)}$ based on \eqref{qj};
\STATE \hspace{0.5cm} Update $\boldsymbol{\psi}^{(j)}$ based on \eqref{psij};
\STATE \hspace{0.5cm} Update $\boldsymbol{\theta}^{(j)}$ based on \eqref{thetaj}
\STATE \hspace{0.5cm} // refined GS
\STATE \hspace{0.5cm} $\mathbf{u}(0) = \{\mathbf{q}^{(j)}, \boldsymbol{\psi}^{(j)}\}$, $t \leftarrow 1$, $\mathcal{E}(0) \leftarrow \emptyset$
\STATE \hspace{0.5cm} \textbf{while} $t < T$:
\STATE \hspace{0.5cm} \hspace{0.5cm} Generate $\mathcal{B}$ and $\mathcal{D}$  and update $\mathbf{u}(t)$ based on \eqref{ui*}
\STATE \hspace{0.5cm} \hspace{0.5cm} Update $\mathcal{E}(t) = \mathcal{E}(t-1) \cup \mathbf{u}(t)$
\STATE \hspace{0.5cm} \hspace{0.5cm} $t \leftarrow t+1$
\STATE \hspace{0.5cm} \textbf{end while}
\STATE \hspace{0.5cm} Update $\{\mathbf{q}^{(j)}, \boldsymbol{\psi}^{(j)} \}$ as  $\{\mathbf{q}^{t^\star}, \boldsymbol{\psi}^{t^\star}\}$ based on \eqref{etaut}
\STATE \hspace{0.5cm} $j \leftarrow j+1$
\STATE \textbf{end while}
\STATE Output: $\mathbf{q}^{(J)}$, $\boldsymbol{\psi}^{(J)}$, $\boldsymbol{\theta}^{(J)}$
\end{algorithmic}
\label{alg1}
\end{algorithm}

Each GS phase commences at the end of each AO iteration after solving the sub-problem (P1.3) for AIRS passive beamforming, consisting of $T$ iterations. Consider its $t$-th iteration and let $\mathbf{u}(t\!-\!1)$ denote the optimized solution of $\mathbf{u}$ in the $(t\!-\!1)$-th iteration, with $\mathbf{u}(t\!-\!1)=\{\mathbf{q}^{t\!-\!1}, \boldsymbol{\psi}^{t\!-\!1} \} = \{ [q_x^{t\!-\!1},q_y^{t\!-\!1},H]^T, [\psi_z^{t\!-\!1},\psi_y^{t\!-\!1},\psi_x^{t\!-\!1}]^T\}$.
In each GS iteration, we select a fixed number of candidate solutions for exploration, denoted as $I$, which is much smaller than the total number of candidate solutions in $\mathcal{S}$. In particular, we calculate the minimum SNRs achievable by the $I$ selected candidate locations based on \eqref{eta79}. Let $\mathbf{u}_i(t)$ denote the $i$-th candidate solution in the $t$-th GS iteration, $i=1,2,\cdots, I$. The $I$ candidate solutions, i.e., $\mathbf{u}_i(t)$'s, are generated as the union of two sets, denoted as $\mathcal{B}$ and $\mathcal{D}$, respectively.
The first set, $\mathcal{B}$, consists of 10 nearby solutions of $\mathbf{u}(t-1)$, including 4 nearby solutions in terms of locations along $x$- and $y$-axes, i.e., 
\begin{align}
\mathbf{u}_1\!(t)\!\!=\!\!\big\{\![q_x^{t\!-\!1}\!\!+\!\Delta_q,q_y^{t\!-\!1}\!\!, \!H]^T\!\!,\boldsymbol{\psi}^{t\!-\!1}\!\big\}, \;
 \mathbf{u}_2\!(t)\!\!=\!\!\big\{\![q_x^{t\!-\!1}\!\!-\!\Delta_q,q_y^{t\!-\!1}\!\!, \!H]^T\!\!,\boldsymbol{\psi}^{t\!-\!1}\!\big\}, \notag \\
\mathbf{u}_3\!(t)\!\!=\!\!\big\{\![q_x^{t\!-\!1}\!\!,q_y^{t\!-\!1}\!\!+\!\Delta_q, \!H]^T\!\!,\boldsymbol{\psi}^{t\!-\!1}\!\big\}, \; 
\mathbf{u}_4\!(t)\!\!=\!\!\big\{\![q_x^{t\!-\!1}\!\!,q_y^{t\!-\!1}\!\!-\!\Delta_q, \!H]^T\!\!,\boldsymbol{\psi}^{t\!-\!1}\!\big\},
\notag 
\end{align}
and 6 nearby solutions in terms of orientations around $x'$-, $y'$-, and $z'$-axes, i.e., 
\begin{align}
\mathbf{u}_5\!(t) \!\!=\!\! \{\mathbf{q}^{t\!-\!1}\!\!\!, [\psi_z^{t\!-\!1}\!\!\!+\!\Delta_\psi,\!\psi_y^{t\!-\!1}\!\!\!,\psi_x^{t\!-\!1}\!]^{\!T} \!\}\!, \;
\mathbf{u}_6\!(t) \!\!=\!\! \{\mathbf{q}^{t\!-\!1}\!\!\!, [\psi_z^{t\!-\!1}\!\!\!\!-\!\Delta_\psi,\!\psi_y^{t\!-\!1}\!\!\!,\psi_x^{t\!-\!1}\!]^{\!T} \!\}\!, \notag \\
\mathbf{u}_7\!(t) \!\!=\!\! \{\mathbf{q}^{t\!-\!1}\!\!\!, [\psi_z^{t\!-\!1}\!\!\!,\psi_y^{t\!-\!1}\!\!\!+\!\Delta_\psi,\!\psi_x^{t\!-\!1}\!]^{\!T} \! \}\!,\;
\mathbf{u}_8\!(t) \!\!=\!\! \{\mathbf{q}^{t\!-\!1}\!\!\!, [\psi_z^{t\!-\!1}\!\!\!,\psi_y^{t\!-\!1}\!\!\!\!-\!\Delta_\psi,\!\psi_x^{t\!-\!1}\!]^{\!T} \! \}\!, \notag \\
\mathbf{u}_9\!(t) \!\!=\!\! \{\mathbf{q}^{t\!-\!1}\!\!\!, [\psi_z^{t\!-\!1}\!\!,\psi_y^{t\!-\!1}\!\!\!,\psi_x^{t\!-\!1}\!\!+\!\Delta_\psi\!]^{\!T} \! \}\!, \,
\mathbf{u}_{10}\!(t) \!\!=\!\! \{\mathbf{q}^{t\!-\!1}\!\!\!, [\psi_z^{t\!-\!1}\!\!\!,\psi_y^{t\!-\!1}\!\!\!,\psi_x^{t\!-\!1}\!\!\!\!-\!\Delta_\psi\!]^{\!T} \! \}\!. \notag 
\end{align}
The second set, $\mathcal{D}$, contains $(I-10)$ solutions randomly selected from the non-selected candidate solutions in the set $\mathcal{S}\setminus\mathcal{B}$.
Note that any solution that is out of the feasible region will be truncated, e.g., $\mathbf{u}_5(t)$ will be truncated as $\{\mathbf{q}^{t\!-\!1}\!\!\!, [\frac{\pi}{2},\psi_y^{t\!-\!1}\!\!\!,\psi_x^{t\!-\!1}]^{\!T} \!\}$ if $\psi_z^{t\!-\!1}\!\!+\!\Delta_\psi > \frac{\pi}{2}$.
In addition, we denote $\mathcal{E}(t-1)$ as the set of all feasible solutions the GS has visited, i.e., $\mathcal{E}(t-1) =\{ \mathbf{u}(1),\;\mathbf{u}(2),\;\cdots,\;\mathbf{u}(t-1) \} $.

The refined GS is achieved by designing a Markov chain for updating the AIRS's location and 3D orientation iteratively, and the transition probability from the solution in the $(t-1)$-th iteration to the $t$-th iteration is given by \cite{GS}
\begin{align}
\text{P}_i^{t} &= \text{Pr}\big\{\mathbf{u}(t)=\mathbf{u}_i(t) | \mathbf{u}(t-1)=\{\mathbf{q}^{t\!-\!1}, \boldsymbol{\psi}^{t\!-\!1} \}\big\}  \notag \\
&=  \frac{e^{\mu \gamma_{\min}(\mathbf{u}_i(t))}}{ \sum_{ \mathbf{u}_i(t) \in \mathcal{B}\cup\mathcal{D}} e^{\mu \gamma_{\min}(\mathbf{u}_i(t))} }, i=1,2,\cdots,I,
\label{stp} 
\end{align}
where $\mu \ge 0$ is a pre-defined scaling parameter. To avoid undesirable bouncing between two solutions with the highest minimum SNR, we manually set $\gamma_{\min}\big(\mathbf{u}_i(t)\big) = \gamma_{\min}\big(\mathbf{u}_i(t)\big)-3$ dB in the case of  $\mathbf{u}_i(t) \in \mathcal{E}(t-1)$. To determine $\mathbf{u}(t)$ based on \eqref{stp}, we randomly generate a float (denoted as $p_{t}$) between 0 and 1, and update 
\begin{equation}
\mathbf{u}(t) = \mathbf{u}_{i^\star}(t), \label{ui*}
\end{equation} 
where $i^{\star}$ is the index satisfying $\sum_{i=1}^{i^\star-1} \text{P}_i^{t} < p_{t} \le \sum_{i=1}^{i^\star} \text{P}_i^{t}$.

At the beginning of the GS phase, the solution is initialized as $\mathbf{u}(0) = \{\mathbf{q}^{(j)}, \boldsymbol{\psi}^{(j)}\}$, and the GS proceeds until the iteration number $t$ reaches a pre-defined maximum number of iterations, denoted by $T$. Finally, among all solutions in $\mathcal{E}(T)$, we choose the solution that yields the maximum minimum SINR as the output of GS, which is given by
\begin{equation}
\mathbf{u}(t^\star) = \{ \mathbf{q}^{t^{\star}}, \boldsymbol{\psi}^{t^{\star}} \} = \arg \max_{\mathbf{u}\in\mathcal{E}(T)}\gamma_{\min}\big(\mathbf{u}\big). \label{etaut}
\end{equation}
The $\{ \mathbf{q}^{(j)},\boldsymbol{\psi}^{(j)}\}$ is then updated as $\{ \mathbf{q}^{t^\star}, \boldsymbol{\psi}^{t^\star}\}$ for the next AO iteration. We summarize the main steps of our proposed enhanced AO with GS in Algorithm 1.

Last, we analyze the complexity of our proposed AO algorithm with GS. 
The complexity order of solving sub-problem (P1.1) via the hybrid search can be expressed as $\mathcal{O}(N_u N_{\text{tot}})+\mathcal{O}(N_u N_{\text{tot}, n^*})$, and that of solving sub-problem (P1.2) is given by $\mathcal{O}(N_u \tilde N_{\text{tot}})+\mathcal{O}(N_u \tilde N_{\text{tot},m^*})$. To solve each sub-problem (P1.3$x$) , it can be shown that the complexity order is given by $\mathcal{O}(N_u^{1.5}N_x^{6.5})$ \cite{For_Complexity}. As such, the complexity of solving (P1.3) is given by $\mathcal{O}(N_u^{1.5}N_x^{6.5})+\mathcal{O}(N_u^{1.5}N_y^{6.5})$. Finally, the complexity order of GS phases per AO iteration is given by $\mathcal{O}(TI)$. It follows that the proposed AO algorithm with GS admits a polynomial computational complexity.

\section{Numerical Results}
In this section, numerical results are provided to evaluate the performance of our proposed UAV-enabled passive 6DMA and validate the theoretical analyses.

\begin{figure}[tb]
\centerline{\includegraphics[width=0.4625\textwidth]{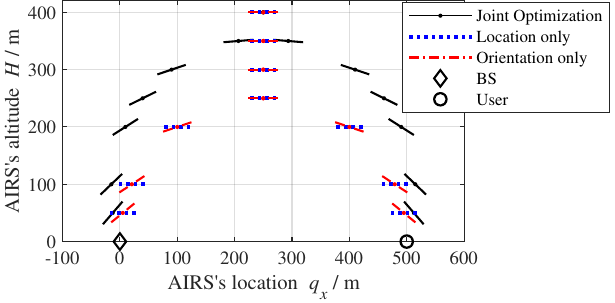}}
\caption{AIRS's orientation versus its location.}
\label{AIRS2}
\end{figure}

\subsection{Simulation Parameters}
 The noise power and the BS's transmit power are set as $ \sigma^2 = -110$ dBm and $P = 20$ dBm, respectively, while the reference path gain is $\beta_0 = -40$ dB. The distance between adjacent transmit antennas at the BS is $d_{tx} = \frac{\lambda}{2}$, and the distance between adjacent reflecting elements of the AIRS is $d_{rs} = \frac{\lambda}{2}$.
The number of the BS's antennas is set to $M = 64$. The number of AIRS reflecting elements per dimension is assumed to be identical as $N_x = N_y = 16$, i.e., $N=256$.
Unless otherwise stated,  the AIRS's altitude is fixed as $H = 100$ m. 
The maximum numbers of AO and GS iterations are $J = 3$ and $T=400$, respectively.
In the GS, the penalty parameter, scaling parameter, number of selected candidate solutions, location interval, and orientation interval are set as $\rho=10$, $\mu=20$, $I=30$, $\Delta_q=5$ m, and $\Delta_\psi=\pi/180$, respectively.  The numbers of sampling points in the coarse- and fine-grained search are $\bar N_x =\bar N_y = N_{x,n^*} = N_{y,n^*} = 100$, $\tilde N_z =\tilde N_y =\tilde N_x = 60$, and $\tilde N_{z,m^*} =\tilde N_{y,m^*} =\tilde N_{x,m^*} = 3$, respectively.

\begin{figure}[tb]
\centerline{\includegraphics[width=0.4525\textwidth]{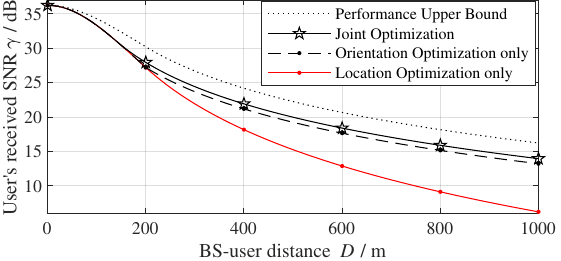}}
\caption{User's received SNR versus BS-user distance.}
\label{SNRvD}
\end{figure}

In the single-user case, the BS-user distance is fixed as $D=500$ m, and the UAV's movement region is set as ${\cal{Q}} =\{q_x|-0.2D\le q_x \le1.2D\}$.
In the multi-user case, we consider two setups of the users' geographic distributions with the number of users fixed as $N_u=3$.
In the first setup, the locations of the users are sparse and given by
$\mathbf{w}_1=[330,240,0]^T$, $\mathbf{w}_2=[650,130,0]^T$, $\mathbf{w}_3=[440,15,0]^T$.
In the second setup, the locations of the users are denser and given by
$\mathbf{w}_1=[655,130,0]^T$, $\mathbf{w}_2=[650,135,0]^T$, $\mathbf{w}_3=[650,130,0]^T$.
The UAV's movement region is set as
$\mathcal{Q} = \{(q_x, q_y) |-140\le q_x \le 790, \; -58 \le q_y \le 298 \}$.

The AIRS's location, orientation, and phase shifts are initialized individually in the AO algorithm based on the following procedures. First, the AIRS's location is initialized as
\begin{equation}
\mathbf{q}^{(1)} = \arg \; \max_{\mathbf{q}\in \mathcal{Q}} \; \min_{l \in \mathcal{N}_u} \frac{1}{\|\mathbf{q} \|^2 \| \mathbf{q}-\mathbf{w}_l\|^2},
\end{equation}
which maximizes the minimum path gain from the BS to all users. Second, the AIRS's orientation is initialized as 
\begin{equation}
\boldsymbol{\psi}^{(1)} = \arg \; \max_{\boldsymbol{\psi}} \min_{l \in \mathcal{N}_u} F_{\text{AG},l}(\mathbf{q}^{(1)}, \boldsymbol{\psi}),
\end{equation}
which maximizes the minimum aperture gain from the BS to all users. Finally, the AIRS's passive beamforming is initialized as $\boldsymbol{\theta}^{(1)}$ by solving (P1.3$x$) and (P1.3$y$) with $\alpha_l = 1$, which maximizes the minimum passive beamforming gain achievable by all users.

\subsection{Single-User Case}
In the single-user case, we consider the following two benchmark schemes, i.e.
\begin{itemize}
\item \!\!AIRS's orientation optimization only with $q_x$ fixed as \eqref{oqx}, which is optimal for isotropic signal reflection.
\item \!\!AIRS's location optimization only with $\psi_y = 0$.
\end{itemize}

First, by varying the AIRS's altitude $H$, Fig.~\ref{AIRS2} depicts the AIRS's optimal orientations and locations under $D\!=\!500$ m by the proposed joint location and orientation optimization and the two benchmark schemes.
It is observed that the AIRS's optimal locations by the two benchmark schemes are identical, which validates our analysis presented at the end of Section III-B-1).
Moreover, it is observed that the AIRS's optimal orientation is symmetric to $q_x\!=\!\frac{D}{2}$, at which the AIRS is parallel to the ground, i.e., no orientation is needed. This observation is consistent with our theoretical analyses provided in Section III-B-2). 
Furthermore, as the AIRS's altitude $H$ decreases, its optimal orientation angle becomes closer to $\frac{\pi}{2}$ or $-\frac{\pi}{2}$, while its optimal location becomes closer to $0$ or $D$. 
On the other hand, as $H$ increases, the AIRS's optimal location and orientation angle approach $q_x=\frac{D}{2}$ and $\psi_y=0$, respectively.
As a result, the AIRS's location and orientation by the joint optimization are mostly different from those by the two benchmark schemes for a moderate $H$. All of the above observations match the theoretical analyses conducted in Section III-B-3) for $H \!\rightarrow \!0$ or $\rightarrow \!\infty$.

\begin{figure}[t]
\centerline{\includegraphics[width=0.4608\textwidth]{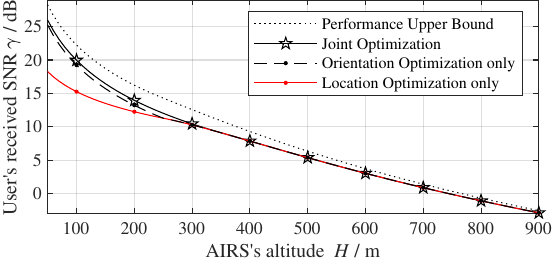}}
\caption{User's received SNR versus AIRS's altitude.}
\label{SNRvH}
\end{figure}

Next, Fig.~\ref{SNRvD} shows the user's received SNR under $H\!=\!100$ m versus the BS-user distance $D$. In addition to the two benchmarks, we also show the user's received SNR in the case of isotropic signal reflection by the AIRS, i.e., $F_{\text{AG},0}(q_x,\psi_y)=1$, which serves as an upper bound on the performance of the considered three schemes. 
It is observed from Fig.~\ref{SNRvD} that as $D$ increases, the SNR performance of all schemes becomes worse, due to the more severe end-to-end path loss. Nonetheless, the proposed joint optimization yields better performance than the other two benchmark schemes. 
Particularly, the AIRS's orientation optimization is observed to significantly outperform its location optimization, which implies that the orientation plays a more significant role than the location in affecting the user's SNR performance. The possible reason is that the effective reflection aperture gain $F_{\text{AG},0}(q_x,\psi_y)$ can rapidly change with the AIRS's orientation angle, which is more dramatic than the change of the end-to-end path loss with the AIRS's location. 
Furthermore, all of the considered schemes are observed to yield comparable SNR performance to the upper bound when $D$ is small, as analyzed in Section III-B-3).

Finally, Fig.~\ref{SNRvH} shows the user's received SNR under $D\!=\!500$ m versus the AIRS's altitude $H$. 
It is first observed from Fig.~\ref{SNRvH} that the SNR performance of all schemes degrades with $\!H\!$ due to the more severe path loss. It is also observed that the joint optimization yields better performance than the other two benchmark schemes for a small $\!H\!$. 
However, as $\!H\!$ increases, the performance gap with them gradually vanishes, and all schemes yield comparable performance to the performance upper bound. This implies that orientation optimization brings lower performance gain for a large $\!H\!$, as analyzed in Section III-B-3).

\begin{figure}[tb]
\centerline{\includegraphics[width=0.4401\textwidth]{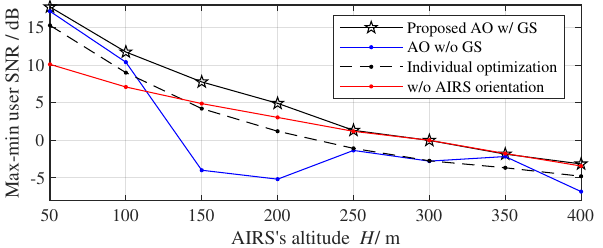}}
\caption{Users' max-min SNR versus AIRS's altitude for sparse user distribution.}
\label{silinorm}
\end{figure} 

\begin{figure}[t]
\centerline{
\includegraphics[width=0.4554\textwidth]{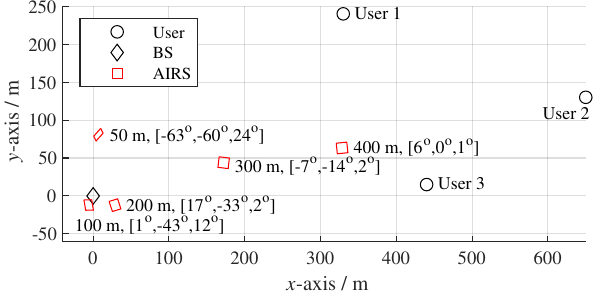}}
\caption{Optimized AIRS position and orientation with different altitudes.}
\label{INTERESTING}
\end{figure} 

\begin{figure*}[h]
\centering
\subfloat[End-to-end path gain]{\includegraphics[width=0.49\textwidth]{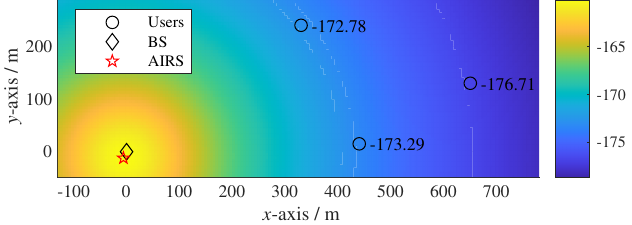}
\label{EPL4}}
\hfil
\subfloat[Effective aperture gain]{\includegraphics[width=0.49\textwidth]{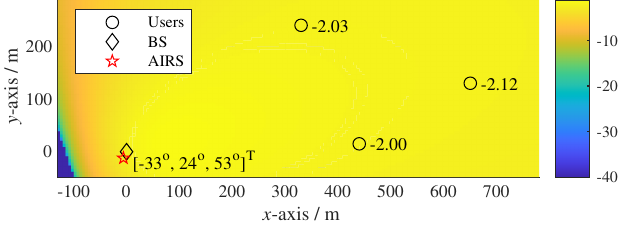}
\label{FAG4}}
\vfil
\subfloat[Passive beamforming gain]{\includegraphics[width=0.49\textwidth]{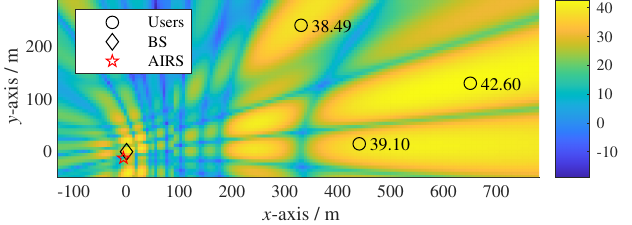}
\label{GBF4}}
\hfil
\subfloat[Max-min user SNR]{\includegraphics[width=0.49\textwidth]{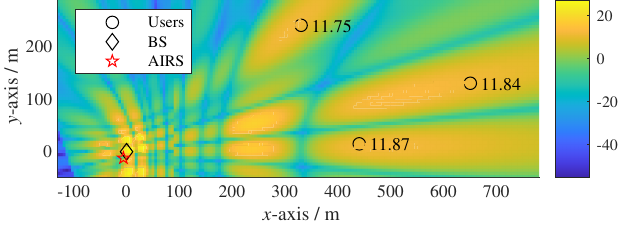}
\label{SNR4}}
\caption{Distribution of the optimized path gain, effective aperture gain, passive beamforming gain, and max-min user SNR over the AIRS's moving region with $H=100$ m.}
\label{pic4}
\end{figure*}

\subsection{Multi-User Case}

In the multi-user case, the proposed enhanced AO algorithm with GS is labeled as ``AO w/ GS'', and we consider the following four benchmark schemes:
\begin{itemize}
\item \!\!AO without GS (AO w/o GS): The conventional AO algorithm described in Section IV-A.
\item \!\!Individual optimization: The AIRS's position, orientation, and phase shifts are individually optimized to maximize the minimum end-to-end path gain, aperture gain, and beamforming gain, respectively, i.e., $\mathbf{q}^{(1)}$, $\boldsymbol{\psi}^{(1)}$ and $\boldsymbol{\theta}^{(1)}$, given at the end of Section V-A.
\item \!\!w/o AIRS orientation: Only the AIRS's location and phase shifts are optimized, while its orientation is fixed as $\boldsymbol{\psi} = [0,0,0]^T$.
\end{itemize}

Fig.~\ref{silinorm}  shows the users' max-min received SNR by different  schemes versus the AIRS's altitude with sparsely distributed users.
It is observed that the SNR performance of all schemes (except the conventional AO without GS) decreases with the AIRS's altitude due to the increased path loss, which is similar to the observation made for the single-user case in Fig.~\ref{SNRvH}. However, the conventional AO without GS is observed to experience significant performance fluctuation as the AIRS's altitude increases. For example, as $H$ increases from 200 m to 250 m, its achieved SNR increases from $-$5 dB to $-$1.3 dB. This implies that it may be trapped by low-quality local optimums and thus result in an unstable SNR performance.
It is also observed that our proposed algorithm outperforms all benchmark schemes considered, thus validating its effectiveness.
However, as the AIRS's altitude $H$ increases, the gap between the proposed scheme and the scheme without AIRS orientation is observed to gradually vanish, implying that the effects of AIRS orientation plays a less significant role. This observation is consistent with the observation made for the single-user case in Fig.~\ref{SNRvH} as well.

Fig.~\ref{INTERESTING} shows the optimized AIRS location and orientation for different altitudes. It is observed that the optimized AIRS's position varies with its altitude. In particular, it is approximately arranged along the line between the BS and user 2 when its altitude is sufficiently high (e.g., $H \ge 100$ m), due to the most severe end-to-end path loss between the BS and user 2 among all users, while its orientation is altered to ensure the effective aperture gain achievable by all users. It is also observed that less orientation is needed for the AIRS with a sufficiently high altitude, due to the less significant role of AIRS orientation in this case, as similarly observed in Figs.~\ref{AIRS2}, \ref{SNRvH}, and \ref{silinorm}.

To gain more insights, Figs.~\ref{pic4}(a)-\ref{pic4}(d) show the distribution of the optimized end-to-end path gain, effective aperture gain, passive beamforming gain and max-min user SNR (all in dB) over the AIRS's moving region $\mathcal{Q}$ by the proposed enhanced AO algorithm, with $H=100$ m.
Note that any effective aperture gain less than $-$40 dB is plotted as $-$40 dB in Fig.~\ref{pic4}(b) due to its excessively small value.
It is observed from Fig.~\ref{pic4}(a)-\ref{pic4}(c) that the proposed algorithm can properly balance all of the path gain, effective aperture gain, and beamforming gain achievable by the three users, thus ensuring the max-min SNR in Fig.~\ref{pic4}(d). 
Compared to the path gain and the effective aperture gain, the fluctuation of the passive beamforming gain is observed to be more significant within $\mathcal{Q}$. Nonetheless, it is worth noting that the maximum passive beamforming gain achievable by each user is given by $10\times\log_{10}(N^2) = 48.2$ dB with $N = 16 ^2$, while the three users are observed to reap a passive beamforming gain of around 40 dB. This implies that the proposed algorithm helps generate multiple high-gain passive beams aligned to each of them. 

\begin{figure}[tb]
\centerline{
\includegraphics[width=0.4402\textwidth]{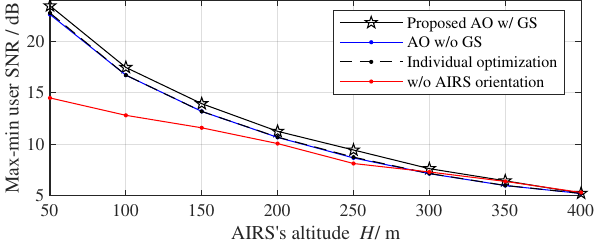}}
\caption{Users' max-min SNR versus AIRS's altitude for dense user distribution.}
\label{silinorm-dense}
\end{figure} 

Last, Fig.~\ref{silinorm-dense} shows the users' max-min received SNR by different schemes versus the AIRS's altitude with densely distributed users.
It is observed that at $H=100$ m, the proposed AO algorithm w/ GS achieves an identical SNR performance to ``Joint Optimization'' at $D=663$ m in Fig.~\ref{SNRvD}, and 663 m is approximately the distance from the BS to user 3. This observation implies that in the case of densely distributed users, the AIRS's location and orientation may suffice to be designed based on a certain user thanks to the small inter-user distances compared to $H$. This fact can also be seen from the smaller performance gap between the proposed algorithm and the benchmark with individual optimization compared to that in Fig.~\ref{silinorm}.

\section {Conclusion}
In this paper, we investigated a joint location, orientation, and beamforming optimization problem for a passive 6DMA-aided multicast system under the practical angle-dependent reflection model. 
In the special single-user case, we unveiled that it suffices to exploit the AIRS's 1D orientation to achieve the optimal performance. Furthermore, we derived the optimal AIRS orientation and location in closed form in some special cases and show their non-trivial relationship with the AIRS's altitude and the BS-user distance.
In the general multi-user case, we proposed an enhanced AO algorithm with GS, where the AIRS's location and orientation were updated iteratively via a probability-based Markov chain to avoid low-quality local optimum.
Numerical results validated our theoretical analyses and demonstrate the superiority of our proposed AO algorithm with GS to other baseline schemes. It was also shown that the AIRS's orientation may have a profound effect on the user SNRs, especially if the AIRS's altitude is not high. Furthermore, the user distribution can affect the efficacy of the joint optimization versus the individual optimization of the AIRS's location and orientation. This paper can be extended to various directions as future work, e.g., the performance optimization of the passive 6DMA for physical-layer security, multi-user broadcasting, non-orthogonal multiple access, etc. It is also interesting to study more general passive 6DMA with tunable relative positions of the AIRS's reflecting elements and evaluate its performance gain.

\appendix
\section*{Proof of Proposition 1}
By noting that only $F$ is affected by $\boldsymbol{\psi}$ in \eqref{finalgamma}, to prove \eqref{prop1}, it is equivalent to prove
\begin {equation} \label{F1}
  F_{\text{AG},0}(q_x, \boldsymbol{\psi}_{\text{2D}}^\star) = F_{\text{AG},0}(q_x,\boldsymbol{\psi}_{\text{3D}}). 
\end{equation}
By substituting \eqref{Fsingle} into \eqref{F1}, it becomes
\begin{align} \label{F1a}
\cos\phi_1(q_x,\boldsymbol{\psi}_{\text{2D}}^\star) &\cos\phi_{2,0}(q_x,\boldsymbol{\psi}_{\text{2D}}^\star) \notag \\
&=\cos\phi_1(q_x,\boldsymbol{\psi}_{\text{3D}}) \cos\phi_{2,0}(q_x,\boldsymbol{\psi}_{\text{3D}}). 
\end{align}

To achieve \eqref{F1a}, we next aim to find a feasible $\boldsymbol{\psi}_{\text{2D}}^\star$ satisfying
\begin{equation} 
\cos\phi_1(q_x,\boldsymbol{\psi}_{\text{2D}}^\star) = \cos\phi_1(q_x,\boldsymbol{\psi}_{\text{3D}}) \label{cosphi1s}
\end{equation}
and 
\begin{equation} \label{cosphi2s}
\cos\phi_{2,0}(q_x,\boldsymbol{\psi}_{\text{2D}}^\star) = \cos\phi_{2,0}(q_x,\boldsymbol{\psi}_{\text{3D}})
\end{equation}
at the same time. To this end, based on \eqref{cosphi1} and \eqref{cosphi2}, we can respectively express the LHSs of \eqref{cosphi1s} and \eqref{cosphi2s} as
\begin {align} \label {lef30}
\cos\!\phi_{1\!}(q_x,\boldsymbol{\psi}_{\text{2D}}^\star)&\!=\! \frac{q_x\!\sin\!\psi_y^\star\cos\!\psi_x^\star \!+\!H\!\cos\!\psi_y^\star \cos\!\psi_x^\star}{\sqrt{q_x^2+H^2}}, \\
\cos\!\phi_{2,0}(q_x,\boldsymbol{\psi}_{\text{2D}}^\star)&\!=\! \frac{(q_x\!\!-\!\!D)\!\sin\!\psi_y^\star\!\cos\!\psi_x^\star\!\!+\!H\!\cos\!\psi_y^\star\!\cos\!\psi_x^\star}{\sqrt{(q_x-D)^2+H^2}},\!\label{lef31} 
\end{align}
with $| \!\sin\!\psi_y^\star\cos\!\psi_x^\star | \!\le\! 1$ and $|\!\cos\!\psi_y^\star\cos\!\psi_x^\star|\!\le \!1$. Similarly, the right hand sides (RHSs) of \eqref{cosphi1s} and \eqref{cosphi2s} can be expressed as
\begin {equation} \label {rig30}
\cos\phi_1(q_x,\boldsymbol{\psi}_{\text{3D}}) = \frac{q_xL_1 +HL_3}{\sqrt{q_x^2+H^2}} ,
\end{equation}
\begin {equation} \label {rig31}
\cos\phi_{2,0}(q_x,\boldsymbol{\psi}_{\text{3D}}) = \frac{(q_x-D)L_1 +HL_3}{\sqrt{(q_x-D)^2+H^2}} ,
\end{equation}
with $L_1 $ and $L_3 $ also satisfying $|L_1 |\le 1$ and $|L_3 |\le 1$.
By comparing \eqref{lef30}-\eqref{lef31} with \eqref{rig30}-\eqref{rig31}, if there exists a set of $\psi_x^\star$ and $\psi_y^\star$ that satisfy both of the following two equations, i.e.,
\begin{align} 
\sin\psi_y^\star\cos\psi_x^\star  &=  L_1 , \label{goal30}  \\
\cos\psi_y^\star\cos\psi_x^\star  &=  L_3 , \label{goal30a} 
\end{align}
then \eqref{cosphi1s} and \eqref{cosphi2s} can be met at the same time.

After some manipulations, it can be shown that  both \eqref{goal30} and \eqref{goal30a} hold if and only if
\begin {equation} \label {rea30}
\cos^2\psi_x^\star = L_1^2+L_3^2.
\end{equation}
Then, if $L_1^2+L_3^2 \le 1$, we can always find a $\psi_x^\star$ that satisfies \eqref{rea30}. Next, we prove
\begin {equation} \label{L12L32}
L_1^2+L_3^2 \le 1.
\end{equation}
To this end, we substitute \eqref{L1}-\eqref{L3} into \eqref{L12L32} and recast it as
\begin{align}  \label{pro1}
2&\sin\psi_z\cos\psi_z\sin\psi_y\sin\psi_x\cos\psi_x + \cos^2\psi_y\cos^2\psi_x  \notag \\
  &+\sin^2\psi_z\sin^2\psi_x + \cos^2\psi_z \sin^2\psi_y\cos^2\psi_x \le 1. 
\end{align}
Note that the RHS of \eqref{pro1}, i.e., 1, can be rewritten as 
\begin{equation} \label{change}
1\!=\!\cos^2\!\psi_z(\cos^2\!\psi_y\!+\sin^2\!\psi_{y\!}) + (\sin^2\!\psi_z \!+ \cos^2\!\psi_{z\!}) \sin^2\!\psi_x.
\end{equation}
By substituting \eqref{change} into \eqref{pro1},  \eqref{pro1} becomes
\begin{align} \label{pro2}
\cos^2\psi_z\sin^2\psi_y\cos^2\psi_x+ 2\sin\psi_z\cos\psi_z\sin\psi_y\sin&\psi_x\cos\psi_x \notag \\
 \le \sin^2\psi_y\cos^2\psi_x+\cos^2\psi_z\sin^2\psi_x.& 
\end{align}
By applying the fact that $\cos^2\!\psi_z\!=\!1\!-\! \sin^2\!\psi_z$ to the first term in \eqref{pro2}, \eqref{pro2} becomes
\begin{align} \label{pro3}
2\sin\psi_z&\cos\psi_z\sin\psi_y\sin\psi_x\cos\psi_x \notag \\
 &\le \sin^2\psi_z\sin^2\psi_y\cos^2\psi_x+\cos^2\psi_z \sin^2 \psi_x. 
\end{align}
Next, we move the LHS of \eqref{pro3} to its RHS and obtain
\begin{equation} \label{pro4}
0 \le  (\sin\psi_z\sin\psi_y\cos\psi_x-\cos\psi_z\sin\psi_x)^2,
\end{equation}
which is always true. Hence, the inequality in \eqref{L12L32} always holds, and we can calculate $\psi_x^\star$ from \eqref{rea30} as
\begin {equation} \label {cospsix}
\psi_x^\star =\pm\arccos \sqrt{L_1^2+L_3^2}.
\end{equation}
By substituting \eqref{cospsix} into \eqref{goal30} and \eqref{goal30a}, we can calculate $\psi_y^\star$ as 
\begin{equation} \label{goal30i}
\psi_y^\star = \arcsin\frac{L_1}{\sqrt{L_1^2 + L_3^2}}.
\end{equation}
The proof of Proposition 1 is thus complete.

\renewcommand{\baselinestretch}{1}
\bibliography{IEEEabrv,mybib}
\bibliographystyle{IEEEtran}

\end{document}